\documentclass[journal]{IEEEtran}
\usepackage{a0size}
\usepackage{amssymb}
\usepackage{multicol}
\usepackage[english]{babel}
\usepackage{epsfig}
\usepackage{bm}
\usepackage{amsfonts,color,amsthm,amsmath, lscape}
\usepackage{url}
\usepackage{algorithm}
\usepackage{algpseudocode}

\usepackage[font=footnotesize]{caption}
\usepackage{subcaption}
\usepackage{epstopdf}
\usepackage{graphicx}
\usepackage{algorithm}
\usepackage{algpseudocode}
\usepackage{cite}
\usepackage{fix2col}
\DeclareMathOperator*{\Maximize}{maximize}
\newtheorem{lemma}{Lemma}
\newtheorem{A_remark}{Remark}
\usepackage{enumitem}
\renewcommand{\figurename}{Fig.}
\addto\captionsenglish{\renewcommand{\figurename}{Fig.}}

   \newcommand{\changet}[1]{{\color{black}#1}}
   \newcommand{\changett}[1]{{\color{black}#1}}

\begin{document}
% Title
\title{{Hybrid Analog and Digital Beamforming for \changet{mmWave OFDM} Large-Scale Antenna Arrays}
% Thanks
%\thanks{Copyright \textcopyright{} 2015 IEEE. Personal use of this material is permitted. However, permission to use this material for any other purposes must be obtained from the IEEE by sending a request to \url{pubs-permissions@ieee.org.}}
\thanks{This work was supported by the
Natural Sciences and Engineering Research Council (NSERC) of Canada.
The materials in this paper have been presented in part at IEEE International Workshop on Signal Processing Advances in Wireless Communications (SPAWC), Edinburgh, UK, July 2016.}
\thanks{The authors are with The Edward S. Rogers Sr. Department of Electrical and Computer Engineering, University of Toronto, 10 King's College
Road, Toronto, Ontario M5S 3G4, Canada (e-mails:
\{fsohrabi, weiyu\}@comm.utoronto.ca).}}
% Authors
\author{Foad~Sohrabi,~\IEEEmembership{Student Member,~IEEE,} and
        Wei~Yu,~\IEEEmembership{Fellow,~IEEE}
}
\maketitle
%%%%%%%%%%%%%%%%%%%%%%%%%%%%%%%%%%%
%%%%%%%%%%%%%%%%%%%%%%%%%%%%%%%%%%%
% Abstract
%%%%%%%%%%%%%%%%%%%%%%%%%%%%%%%%%%%
\begin{abstract}
Hybrid analog and digital beamforming is a promising candidate for large-scale millimeter wave (mmWave) multiple-input multiple-output (MIMO) systems 
because of its ability to
significantly reduce the hardware complexity of the conventional fully-digital beamforming schemes while being capable of approaching the performance of fully-digital schemes. Most of the prior work on hybrid beamforming considers narrowband channels. However, broadband systems such as mmWave systems are frequency-selective. In broadband systems, it is desirable to design common analog beamformer for the entire band while employing different digital (baseband) beamformers in different frequency sub-bands. This paper considers the hybrid beamforming design for systems with orthogonal frequency division multiplexing (OFDM) modulation. First, for a single-user MIMO (SU-MIMO) system where the hybrid beamforming architecture is employed at both transmitter and receiver, we show that hybrid beamforming with a small number of radio frequency (RF) chains can asymptotically approach the performance of fully-digital beamforming for \changett{a} sufficiently large number of transceiver antennas due to the sparse nature of the mmWave channels. For systems with \changett{a} practical number of antennas, we then propose a unified heuristic design for two different hybrid beamforming structures, the fully-connected and the partially-connected structures, to maximize the overall spectral efficiency of a mmWave MIMO system. \changet{Numerical results are provided to show that the proposed algorithm outperforms the existing hybrid beamforming methods and for the fully-connected architecture the proposed algorithm can achieve spectral efficiency very close to that of the optimal fully-digital beamforming \changett{but with much fewer RF chains.} Second, for the multiuser multiple-input single-output (MU-MISO) case, we propose a heuristic hybrid percoding design to maximize the weighted sum rate in the downlink and show numerically that the proposed algorithm with practical number of RF chains can already approach the performance of fully-digital beamforming.}
\end{abstract}
%%%%%%%%%%%%%%%%%%%%%%%%%%%%%%%%%%%
% Abstract
%%%%%%%%%%%%%%%%%%%%%%%%%%%%%%%%%%%
%\begin{IEEEkeywords}
%IEEE, IEEEtran, journal, \LaTeX, paper, template.
%\end{IEEEkeywords}
%\IEEEpeerreviewmaketitle
%%%%%%%%%%%%%%%%%%%%%%%%%%%%%%%%%%%
%Introduction
%%%%%%%%%%%%%%%%%%%%%%%%%%%%%%%%%%%
\section{Introduction}
Millimeter wave communication is a promising candidate that can address the challenge of bandwidth shortage 
for the next generation of wireless cellular communication systems \cite{yong2006overview,pi2011introduction,rappaport2013millimeter,rangan2014millimeter}. Although pathloss and absorption are more severe in mmWave frequencies as compared to the conventional frequency bands, more antennas can be packed within a relatively small physical dimension in mmWave frequencies. This leads to the use of large-scale antenna arrays at the transceivers in mmWave communication systems
which can potentially overcome the poor propagation characteristics of the channel \cite{haider2014cellular,doan2004design,wang2015multi}.  One of the major challenges in designing transceivers for mmWave system using large-scale antenna arrays is that the implementation of the conventional fully-digital beamforming schemes such as in \cite{telatar1999capacity,shi2011iteratively,wiesel2008zero} may not be practical because fully-digital beamforming schemes require one separate RF chain for each antenna element. This leads to high hardware complexity and excessive power consumption. 

To address the hardware limitation of fully-digital beamforming, this paper adopts the analog and digital hybrid beamforming architecture in which the overall beamformer consists of a low-dimensional digital beamformer and a high-dimensional analog beamformer implemented using a network of simple analog phase shifters \cite{zhang2005variable,el2013spatially,chinawei,liang2014low,sohrabi2016hybrid,alkhateeb2013hybrid,Wei_Foad_SPAWC2015}.

 The idea of hybrid analog and digital beamforming is first presented by \cite{zhang2005variable} under the name of soft antenna selection for single-carrier SU-MIMO systems. The idea is reintroduced for mmWave frequency spectrum in \cite{el2013spatially}. In particular, a hybrid precoding algorithm for single-carrier SU-MIMO systems is presented in \cite{el2013spatially} by exploiting the sparse nature of mmWave channels. It is shown in \cite{sohrabi2016hybrid} that for flat-fading channels, if the number of RF chains at the transceivers is twice the number of data streams, hybrid beamforming can achieve the full capacity of the MIMO channel. Further, hybrid precoding design problem for a MU-MISO system is considered in \cite{liang2014low,sohrabi2016hybrid}. The common message from different algorithms presented in \cite{zhang2005variable,el2013spatially,chinawei,liang2014low,sohrabi2016hybrid} is that  hybrid beamforming with the number of RF chains in the order of number of data streams can approach the performance of fully-digital beamforming baselines in flat-fading channels.

Most of the aforementioned results are restricted to the flat-fading channels. However, mmWave systems are expected to operate on broadband channels with frequency selectivity. The main challenge in designing hybrid beamforming for frequency-selective channels is that of how to design a common analog beamformer shared across all the subcarriers while adopting digital beamforming weights on a per-subcarrier basis. This important feature differentiates hybrid beamforming design in frequency-selective channels from that in flat-fading channels and motivates us to consider the hybrid beamforming design for mmWave systems with OFDM modulation.

\subsection{Main Contributions}
This paper first considers the beamforming design in a mmWave SU-MIMO system in which both transmitter and receiver employ large-scale antenna arrays with hybrid beamforming structure. \changet{In such a system, we show that with \changett{a} sufficiently large number of antennas the covariance matrices of the channel at different subcarriers are approximately the same and hence they share approximately the same set of eigenvectors.} This asymptotic feature, which is due to the sparsity of the mmWave channels, suggests that the optimal fully-digital eigen-beamformers at all subcarriers are approximately the same. Based on this property, this paper proposes a hybrid beamforming design that can asymptotically realize the optimal fully-digital eigen-beamforming. Although this result is valid only for extremely large number of antennas, it provides intuition as to why hybrid beamforming can approach the performance of the fully-digital beamforming in 
broadband frequency-selective 
mmWave channels.
 
 This paper also considers hybrid beamforming design for practical size of antenna arrays, e.g., arrays with $32$-$128$ antennas, in a typical mmWave propagation environment. In particular, we propose a unified heuristic algorithm for beamforming design for two different hybrid architectures, the fully-connected and partially-connected architectures, to maximize the overall rate under power spectral density constraint in each sub-band.
 In order to understand the design limits of the architecture, this paper makes the simplifying assumption 
 that perfect channel state information (CSI) is available.
Towards developing this algorithm, we show that it is possible to transform the analog precoding design problem for frequency-selective channels into an analog precoding design problem for flat-fading channels in which the covariance matrix is given by the average of the covariance matrices of frequency domain channels, i.e., $\frac{1}{K} \sum_k \mathbf{H}[k]^H\mathbf{H}[k]$, where $\mathbf{H}[k]$ is the channel at $k^\text{th}$ subcarrier and $K$ is the total number of subcarriers. This transformation enables us to employ the analog beamforming design algorithm already proposed in \cite{sohrabi2016hybrid} for single-carrier systems with flat-fading channel model.  Finally, we find the optimal closed-form solution for the digital beamformers at each sub-carrier for the already designed analog beamformers.

This paper further considers the multiuser case, more specifically, the hybrid precoding design problem in a mmWave MU-MISO system to maximize the downlink weighted sum rate under a power constraint per subcarrier. This problem differs from spectral efficiency maximization in a SU-MIMO system in two respects. First, in the MU-MISO scenario the receiving antennas are not collocated, therefore the beamforming  design  for MU-MISO  case should consider the effect of inter-user interference. Second, the priority weights of the data streams of different users may be unequal in a MU-MISO system, while different data streams in a SU-MIMO system always have the same priority weights. In order to tackle this problem, we propose the following simple design strategy. First, the analog precoder is designed based on the algorithm developed for SU-MIMO scenario assuming that the users are cooperative and have equal priority weights. Further, when the analog precoder is fixed, iterative weighted minimum mean squared error (WMMSE) approach is employed to design the digital precoders at each subcarrier to deal with the inter-user interference and also the different priority of different data streams. Numerical results show that the fully-connected hybrid architecture with this design can already approach the performance of the fully-digital WMMSE beamforming.

\subsection{Related Work}
Several recent works have considered the use of hybrid beamforming architecture for frequency-selective channels \cite{alkhateeb2015frequency,yu2016alternating,park2016dynamic,kong2015wideband,zhou2016channel}. In \cite{alkhateeb2015frequency}, the problem of hybrid beamforming design for maximizing the spectral efficiency in a SU-MIMO mmWave system with limited feedback is considered. 
The authors of \cite{alkhateeb2015frequency} first develop a hybrid analog-digital codebook design scheme for broadband mmWave systems, then propose a hybrid precoding algorithm for the given codebook based on Gram-Schmidt orthogonalization. 

The paper \cite{yu2016alternating} also considers the spectral efficiency maximization problem for a SU-MIMO system, but unlike \cite{alkhateeb2015frequency} the beamformers are not restricted to come from a fixed codebook. The algorithm in 
\cite{yu2016alternating} seeks to minimize the norm distance between the optimal fully-digital beamformers and the overall hybrid beamformers instead of tackling the original problem of spectral efficiency maximization directly.

In \cite{park2016dynamic}, a heuristic algorithm is devised to design the hybrid precoders to maximize the overall rate for a SU-MIMO system in which hybrid architecture is only employed at the transmitter. Taking different approach as compared to this paper, it is also shown in \cite{park2016dynamic} that the average of the covariance matrices of frequency domain channels is an important metric in designing the analog precoder.
 
 Through numerical simulations, we show that under typical parameter settings the proposed algorithm for SU-MIMO scenario achieves a better performance as compared to the algorithms in \cite{yu2016alternating} and \cite{park2016dynamic}.

The weighted sum rate maximization problem under the total power constraint for the downlink of OFDM-based MU-MISO systems is recently considered in \cite{kong2015wideband}. The authors in \cite{kong2015wideband} devise an alternating optimization algorithm based on the equivalence between the sum rate maximization problem and the weighted sum mean square error (MSE) minimization. That algorithm is only applicable to the setting where the total power constraint is considered. \changet{ However, in practice for wideband systems it is desirable to design the precoders such that the per subcarrier power constraint; i.e. power spectrum density constraint, is satisfied \cite{cheung2007adaptive,gueguen2007ofdm}}. In contrast to \cite{kong2015wideband}, the proposed algorithm in this paper addresses the hybrid precoding design problem for per subcarrier power constraint.  

\changet{Throughout this paper, we assume the availability of the prefect CSI. This assumption 
is made in order to understand the capability of the hybrid architecture, but it can also be reasonable 
in certain cases.
In \cite{zhou2016channel}, it is shown that the channel coefficients can be estimated accurately by exploiting the intrinsic low-rank structure of the mmWave channels.}
\subsection{Paper Organization and Notations}
The remainder of this paper is organized as follows. Section~\ref{sec:model} introduces the system model and the problem formulation for a SU-MIMO system. Section~\ref{sec_asy} considers asymptotic analysis of hybrid beamforming in a SU-MIMO system, while Section~\ref{sec_design} considers hybrid beamforming design for practical number of antennas.  Section~\ref{sec:miso} is devoted to present the system model and hybrid precoding design algorithm for a MU-MISO system. Simulation results are provided in Section~\ref{sec_sim} and conclusions are drawn in Section~\ref{sec:con}.

This paper uses lower-case letters for scalars, lower-case bold face letters for vectors and upper-case bold face letters for matrices. The real part of a complex scalar $s$ are denoted by $\operatorname{Re}\{s\}$.  For a matrix $\mathbf{A}$, the element in the $i^\text{th}$ row and the $j^\text{th}$ column is denoted by $\mathbf{A}(i,j)$. Further, we use the superscript ${}^H$ to denote the Hermitian transpose of a matrix
and superscript ${}^*$ to denote the complex conjugate.
 The identity
matrix with appropriate dimensions is denoted by $\mathbf{I}$; $\mathbb{C}^{m\times n}$ denotes an $m$ by $n$ dimensional complex space; $\mathcal{CN}(\mathbf{0},\mathbf{R})$ represents the zero-mean complex Gaussian distribution with covariance matrix $\mathbf{R}$.
Further, the notations $\operatorname{Tr}(\cdot)$, $\operatorname{log}_2(\cdot)$ and $\mathbb{E}[\cdot] $ represent the
trace, binary logarithm and expectation operators, respectively; $|\cdot|$ represent determinant or absolute value depending on context. Finally, for a column vector $\mathbf{v}$, $\operatorname{diag}(\cdot)$ returns a diagonal matrix with elements of  $\mathbf{v}$ as the diagonal elements.
%% Fig1
\begin{figure*}[t]
        \centering
        \includegraphics[width=0.9\textwidth]{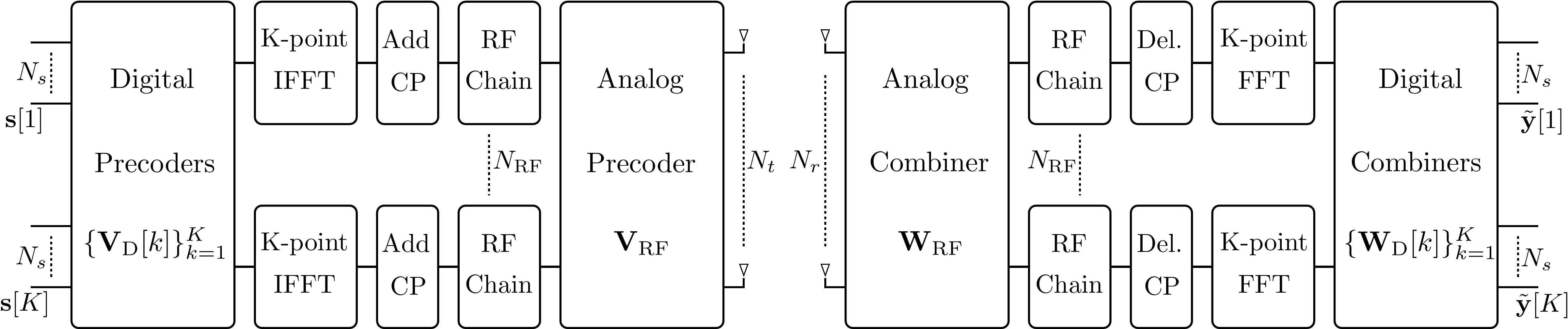}
        \caption{ {A block diagram of an OFDM-based large-scale MIMO system with hybrid analog and digital beamforming architecture at the transceivers.}}
        \label{fig:sys_model}
\end{figure*}

%%%%%%%%%%%%%%%%%%%%%%%%%%%%%%%%%%%
%System Model
%%%%%%%%%%%%%%%%%%%%%%%%%%%%%%%%%%%
\section{System Model for SU-MIMO}
\label{sec:model}
We begin by treating the hybrid beamforming design problem for the single-user frequency-selective channel.
Consider an OFDM-based large-scale MIMO system in which a transmitter equipped with $N_t$ antennas serves a receiver equipped with $N_r$ antennas by sending $N_s$ data symbols per frequency tone. \changet{In general the number of data symbols can be different for different frequency tones, 
\changett{ however for simplicity, this paper restricts attention to the case with an equal number of data streams for all subcarriers, since for mmWave systems with highly correlated channels, all the subchannels are typically low rank.}} Further, in practical large-scale MIMO systems, the number of available RF chains\footnote{\changet{To simplify the notation, we  assume that the number of RF chains at the transmitter and the receiver is identical, however the  results can be easily applied to the general setting.} }, $N_\text{RF}$, is typically much smaller than the number of transceiver antennas, i.e., $N_\text{RF} \ll \min(N_t,N_r)$. This prohibits the implementation of conventional fully-digital beamforming methods which require one RF chain per antenna element. In this paper, we adopt the hybrid analog-digital beamforming architecture, shown in Fig.~\ref{fig:sys_model}, to address this hardware limitation challenge. In the hybrid beamforming architecture, the overall beamformer consists of a low-dimensional digital (baseband) beamformer and a high-dimensional analog (RF) beamformer implemented using simple analog components.

 %%%%%%%%%%%%%%% Subsection 2-A
\subsection{Signal Model in Hybrid Beamforming}
In \changett{the} OFDM-based hybrid beamforming architecture shown in Fig.~\ref{fig:sys_model}, the transmitter first precodes $N_s$ data symbols $\mathbf{s}[k]$ at each subcarrier $k=1,\dots,K$, using a low-dimensional digital precoder, $\mathbf{V}_\text{D}[k] \in \mathbb{C}^{N_\text{RF} \times N_s}$, then transforms the signals to the time domain by using $N_\text{RF}$ $K$-point inverse fast Fourier transforms (IFFTs). After adding \changett{cyclic prefixes}, the transmitter employs an analog precoding matrix $\mathbf{V}_\text{
RF} \in \mathbb{C}^{N_t \times N_\text{RF}}$, to generate the final transmitted signal. Since the analog precoder is a post-IFFT module, the analog precoder is identical for all subcarriers. This is the key challenge in designing the hybrid beamformers in OFDM systems as compared to single-carrier systems. By this consideration, the final transmitted signal at subcarrier $k$ is
%1
\begin{equation}
\mathbf{x}[k] = \mathbf{V_\text{RF}} \mathbf{V_\text{D}}[k] \mathbf{s}[k], 
\end{equation}
where $\mathbf{s}[k] \in \mathbb{C}^{N_s \times 1}$ is the vector of transmitted data symbols at subcarrier $k$ with $\mathbb{E}\{\mathbf{s}[k]\mathbf{s}[k]^H\} = \mathbf{I}_{N_s}$. Assuming a block-fading channel model, the received signal at subcarrier $k$ is
%2
\begin{equation}
\mathbf{y}[k] = \mathbf{H}[k] \mathbf{V_\text{RF}} \mathbf{V_\text{D}}[k] \mathbf{s}[k] + \mathbf{z}[k], 
\end{equation}
where $\mathbf{H}[k] \in \mathbb{C}^{N_r\times N_t}$ and $\mathbf{z}[k] \sim \mathcal{CN}(\mathbf{0},\sigma^2 \mathbf{I}_{N_r}) $ are the channel matrix and additive white Gaussian noise for subcarrier $k$, respectively.

At the receiver, 
the received signals of all subcarriers are initially processed using an analog combiner, $\mathbf{W}_\text{RF} \in \mathbb{C}^{N_r \times N_\text{RF}}$. Then, the cyclic prefix is removed and $N_\text{RF}$ $K$-point fast Fourier transforms (FFTs) are applied to recover the frequency domain signals. Finally, by employing a low-dimensional digital combiner per subcarrier, $\mathbf{W}_\text{D}[k] \in \mathbb{C}^{N_\text{RF} \times N_s}$, the receiver obtains the final processed signal as
%3
\begin{equation}
\tilde{\mathbf{y}} [k] = {\mathbf{W}_\text{t}[k]}^H   \mathbf{H}[k] \mathbf{V}_\text{t}[k]  \mathbf{s}[k] + {\mathbf{W}_\text{t}[k]}^H  \mathbf{z}[k],
\end{equation}
in which $\mathbf{V}_\text{t}[k] =  \mathbf{V}_\text{RF}\mathbf{V}_\text{D}[k]$ and $\mathbf{W}_\text{t}[k] =  \mathbf{W}_\text{RF}\mathbf{W}_\text{D}[k]$ are the overall hybrid precoder and combiner for the $k^\text{th}$ subcarrier, respectively.

%% Fig2
\begin{figure}
    \centering
    \begin{subfigure}[b]{.21\textwidth}
        \includegraphics[width=\textwidth]{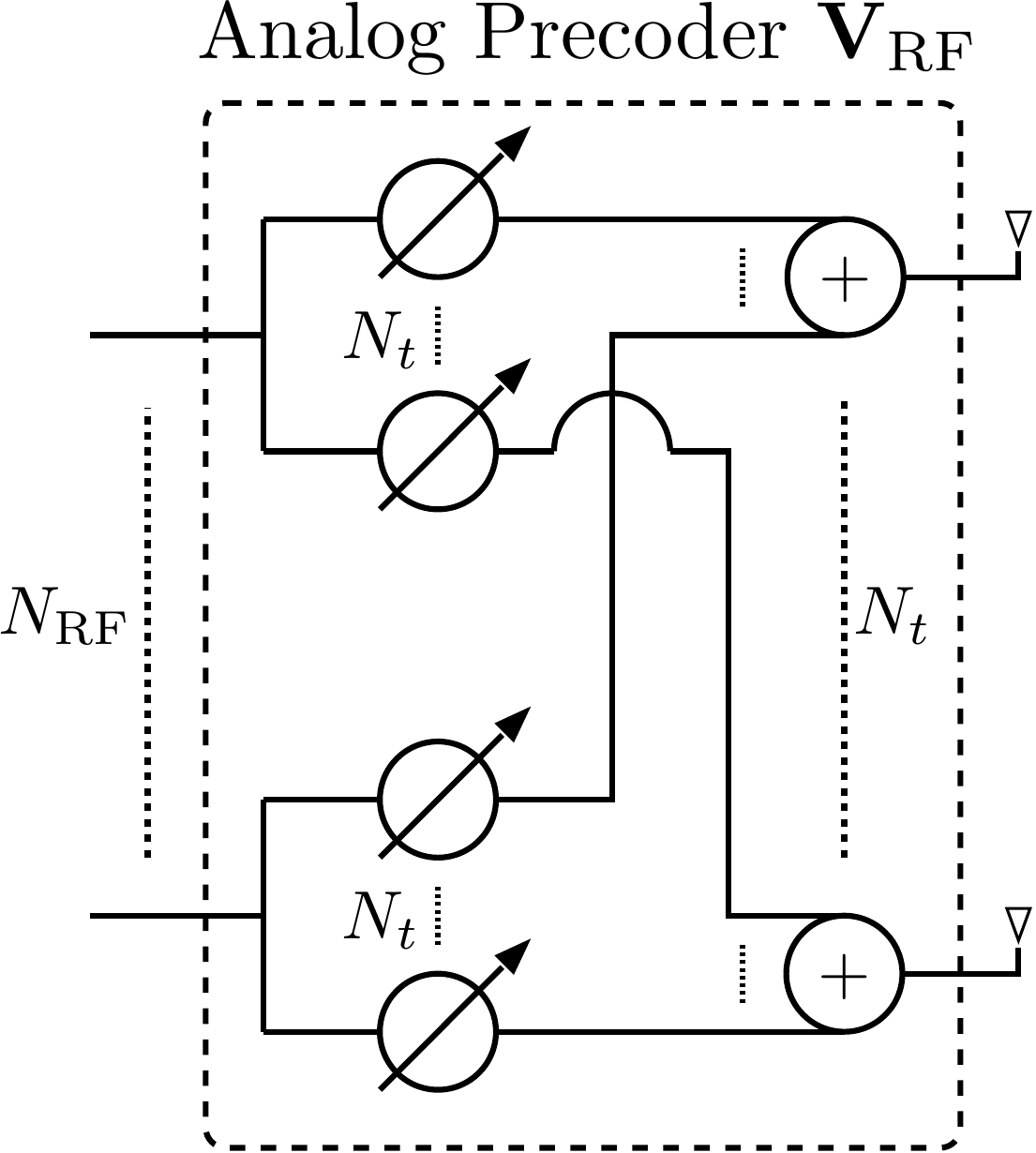}
        \caption{}
        \label{fig:fully}
    \end{subfigure}
    ~ 
    \begin{subfigure}[b]{0.2\textwidth}
        \includegraphics[width=\textwidth]{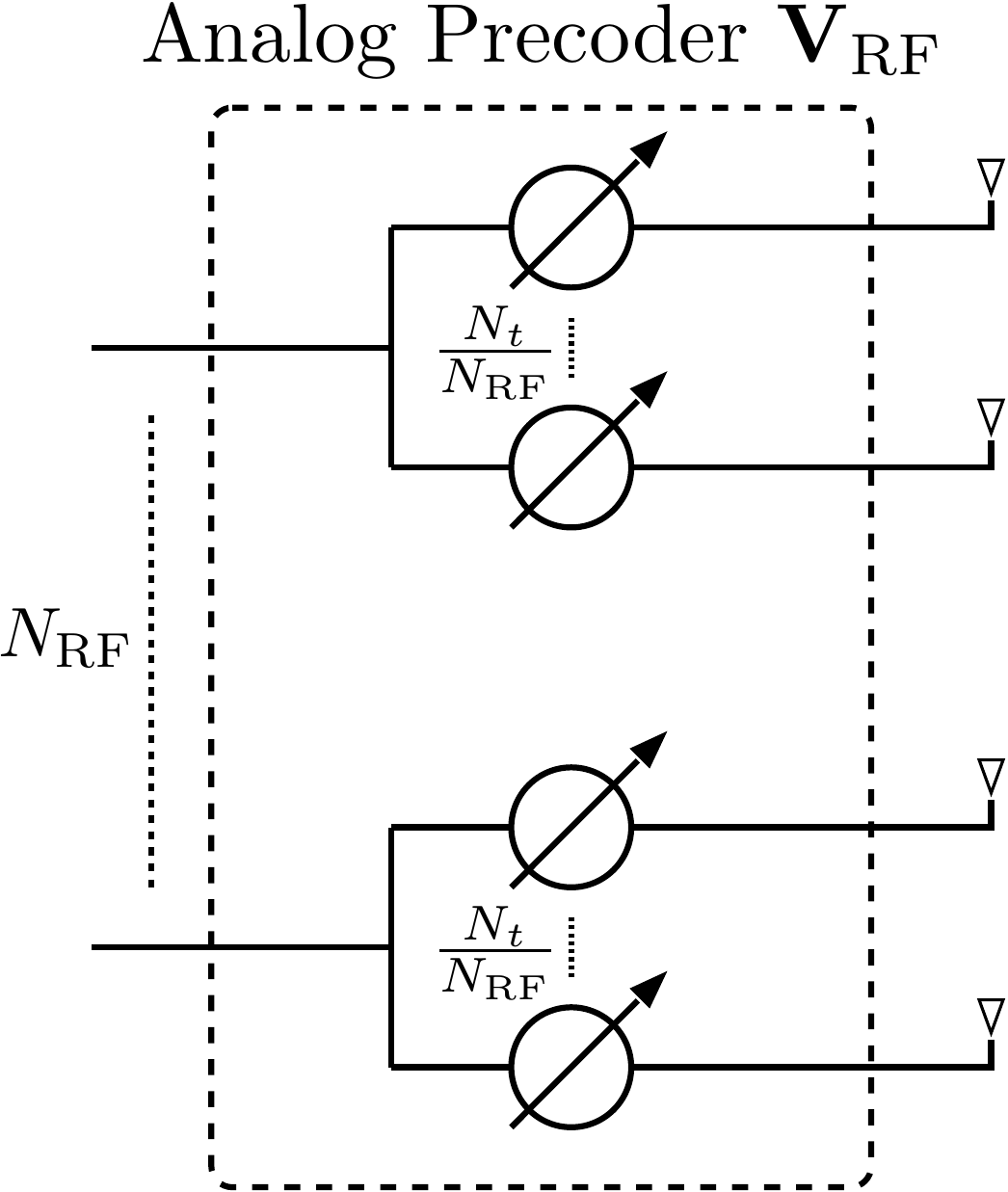}
        \caption{}
        \label{fig:partially}
    \end{subfigure}
    \caption{(a) The architecture of an analog precoder with fully-connected structure. (b) The architecture of an analog precoder with partially-connected structure. }\label{fig:system_models}
\end{figure}
 %%%%%%%%%%%%%%% Subsection 2-B 
 \subsection{Structure of Analog Beamformer}
The analog part of the hybrid beamformer is typically implemented using simple analog components such as analog adders and analog phase shifters which can only change the phase of signals. This results in some constraints on the analog beamforming matrix depending on the structure of the analog beamformer. In this paper, we focus on two widely used analog beamforming structures: the fully-connected and the partially-connected structures.

\textbf{Fully-connected Architecture:} In this structure, \changett{each RF chain is connected} to all the antenna elements via a network of phase shifters as shown in 
%Fig.~\ref{fig:fully}. 
Fig.~2(a).
This results in constant modulus norm constraint on all elements of the analog beamforming matrices, i.e., $|\mathbf{V}_\text{RF}(i,j)| = |\mathbf{W}_\text{RF}(i,j)| = 1, \forall i, j$. Further, as it can be seen in 
%Fig.~\ref{fig:fully}, 
Fig.~2(a),
the total number of phase shifters in this architecture is $N_tN_\text{RF}$.
 
\textbf{Partially-connected Architecture:} Unlike the fully-connected structure, in the partially-connected structure, \changett{each RF chain is connected only to} one sub-array with $N_t/N_\text{RF}$ antennas as shown in Fig.~2(b). Therefore, the analog beamforming matrix in the partially-connected architecture has a block diagonal format, i.e., the analog precoder has the form,
%4
\begin{equation}
\label{partially_connected_structure}
\mathbf{V}_\text{RF} = 
\begin{bmatrix} 
\mathbf{v}_1 & \mathbf{0} & \cdots & \mathbf{0}\\
\mathbf{0} & \mathbf{v}_2 &  & \mathbf{0}\\
\mathbf{0} & \mathbf{0} & \ddots & \mathbf{0}\\
\mathbf{0} & \mathbf{0} & \cdots & \mathbf{v}_{N_\text{RF}} 
\end{bmatrix},
\end{equation}
where each element of the vector $\mathbf{v}_i$ satisfies the constant modulus constraint. The total number of phase shifters in this structure is $N_t$ which means that the hardware complexity of the RF beamformer is reduced by a factor of $N_\text{RF}$.

It is shown in \cite{yu2016alternating} that there is a performance-complexity trade-off in choosing the above structures.
\changett{The fully-connected structure can achieve full beamforming gain with full phase control, while the partially-connect structure has limited phase control, thus cannot achieve full beamforming gain in all cases.} On the other hand, the hardware implementation complexity and power consumption of the partially-connected architecture are much lower as compared to those of the fully-connected structure. This paper seeks to propose a general design algorithm which can handle the design of hybrid beamforming for both analog beamformer structures.
 %%%%%%%%%%%%%%% Subsection 2-C
\subsection{Problem Formulation}
The problem of interest for SU-MIMO case is to design the hybrid analog and digital beamformers under either fully or partially connected structures in order to
 maximize the overall spectral efficiency of the system under a power spectral density constraint for each subcarrier. We assume the availability of perfect CSI. In this case, the problem can be formulated as 
%5
\begin{subequations}
\begin{align}
\label{main_problem}
\displaystyle{\Maximize_{\mathbf{V}_\text{RF}, \mathbf{W}_\text{RF},\{\mathbf{V}_\text{D}[k],\mathbf{W}_\text{D}[k] \}_{k=1}^{K} }}  &~  \frac{1}{K} \sum_{k=1}^{K} R[k]\\ 
\text{subject to}  \quad\quad\quad &~ \operatorname{Tr}(\mathbf{V}_\text{t}[k] {\mathbf{V}_\text{t}[k]}^{H}) \leq P,\\ 
\label{4c}
&~|\mathbf{V}_\text{RF}(i,j)|  = 1,  \forall (i,j) \in \mathcal{F}_t,\\
&~|\mathbf{W}_\text{RF}(i,j)| = 1, \forall (i,j) \in \mathcal{F}_r,
\end{align}
\end{subequations}
where $P$ is the total transmit power budget per subcarrier, $\mathcal{F}_t$ ($\mathcal{F}_r$) is the set of non-zero elements of analog precoder (combiner), and $R[k]$ is the overall spectral efficiency of the subcarrier $k$ which for Gaussian signalling is 
%6
\begin{equation}
\label{rate}
R[k] = \log_2 \Bigl| \mathbf{I}_{N_r}  + \frac{1}{\sigma^2}  \mathbf{C}[k] \mathbf{H}[k] \mathbf{V}_\text{t}[k] \mathbf{V}_\text{t}[k]^{H} {\mathbf{H}[k]}^{H}\Bigr|,
\end{equation}
in which 
$\mathbf{C}[k] = \mathbf{W}_\text{t}[k]({\mathbf{W}_\text{t}[k]}^H \mathbf{W}_\text{t}[k])^{-1} {\mathbf{W}_\text{t}[k]}^H$, $\mathbf{V}_\text{t}[k] =  \mathbf{V}_\text{RF}\mathbf{V}_\text{D}[k]$ and $\mathbf{W}_\text{t}[k] =  \mathbf{W}_\text{RF}\mathbf{W}_\text{D}[k]$.
 %%%%%%%%%%%%%%% Subsection 2-D
 \subsection{Channel Model}
 \label{subsec_ch_model}
It is known that the channel of a mmWave propagation environment does not follow the conventional rich-scattering model because the number of scatterers in such an environment is limited \cite{pi2011introduction}. In fact, the mmWave propagation environment is typically modeled as a geometric channel with $N_\text{c}$ scattering clusters and $N_\text{sc}$ scatterers within each cluster. In this model, the channel matrix of subcarrier $k$ is given as\cite{yu2016alternating}
%7
\begin{equation}
\label{eq_ch_model}
\mathbf{H}[k] =  \sum_{c=1}^{N_\text{c}}\sum_{\ell=1}^{N_\text{sc}} \alpha_{c\ell} \mathbf{a}_r(\phi_{c\ell}^r) \mathbf{a}_t(\phi_{c\ell}^t)^H e^{-j2\pi \psi_c \frac{k}{K}},
\end{equation}
where $\psi_c$ is \changett{proportional} to the phase shift in $c^\text{th}$ scattering cluster and $\alpha_{c\ell} \sim \mathcal{CN}(0,{\frac{N_tN_r}{N_\text{c}N_\text{sc}}})$, $\phi_{c\ell}^r$ and $\phi_{c\ell}^t$ are the scaled complex gain,  angles of arrival and departure for the $\ell^\text{th}$ path in the $c^\text{th}$ cluster, respectively. Further, $\mathbf{a}_r(\cdot)$ and $\mathbf{a}_t(\cdot)$ are the antenna array response vectors for the receiver and the transmitter, respectively. The antenna array response vectors $\mathbf{a}_r(\cdot)$ and $\mathbf{a}_t(\cdot)$ are functions of the antenna array structure and each element of these vectors typically satisfies a constant modulus norm constraint. For instant, the antenna array response vector in a uniform linear array configuration with $N$ antennas and antenna spacing of $d$ is modeled as
%8 
\begin{equation}
\mathbf{a}(\phi) = \frac{1}{\sqrt{N}} [1,e^{j\frac{2\pi}{\lambda}d\sin(\phi)}, \dots,e^{j\frac{2\pi}{\lambda}(N-1)d\sin(\phi)}]^T,
\end{equation}
where $\lambda$ is the signal wavelength. Note that in this representation it is assumed that the operating frequency is much larger than the total bandwidth of the OFDM system such that the signal wavelength in all subcarriers can  approximately be considered equal. 

The channel matrix in \eqref{eq_ch_model} can also be written in a more compact form as 
%9
 \begin{equation}
 \label{eq_ch_compact}
 \mathbf{H}[k] = \mathbf{A}_r \operatorname{diag}({\boldsymbol{\alpha}}[k]) \mathbf{A}_t^H,
 \end{equation}
 where
 \begin{eqnarray} \nonumber
 \mathbf{A}_r  &=& [\mathbf{a}_r(\phi_{11}^r),\mathbf{a}_r(\phi_{12}^r),\dots,\mathbf{a}_r(\phi_{N_\text{c}N_\text{sc}}^r)],\\ \nonumber
 \mathbf{A}_t  &=& [\mathbf{a}_t(\phi_{11}^t),\mathbf{a}_t(\phi_{12}^t),\dots,\mathbf{a}_t(\phi_{N_\text{c}N_\text{sc}}^t)],\\ \nonumber
 \boldsymbol{\alpha}[k] &=& [\alpha_{11}e^{-j\theta_{1}[k]},\alpha_{12}e^{-j\theta_{1}[k]},\dots,\alpha_{N_\text{c}N_\text{sc}}e^{-j\theta_{N_\text{c}}[k]}],
 \end{eqnarray}
 and $\theta_{c}[k] =  \frac{2\pi \psi_c k}{K}$.
%%%%%%%%%%%%%%%%%%%%%%%%%%%%%%%%%%%
%  Asymptotic Beamforming Design
%%%%%%%%%%%%%%%%%%%%%%%%%%%%%%%%%%%
 \section{Asymptotic Beamforming Design for SU-MIMO}
 \label{sec_asy}
 In this section, we exploit the sparse nature of the frequency-selective mmWave propagation environment introduced in Section~\ref{subsec_ch_model} to show that for a fixed number of data streams, $N_s$, hybrid beamforming architecture with only $N_s$ RF chains, i.e., $N_\text{RF} = N_s$, can asymptotically realize the optimal fully-digital beamformer when $N_t, N_r \to \infty$.  Note that this result is known in the hybrid beamforming literature for single-carrier systems \cite{el2013spatially,el2012capacity}. In this section, we generalize that result to OFDM-based SU-MIMO systems with frequency-selective channels to explain why the hybrid beamforming is expected to work well in mmWave frequency-selective channels
 despite being much less complex than fully-digital beamforming.

It is well-known that the optimal linear fully-digital precoder in each subcarrier, $\mathbf{V}_\text{opt}[k] \in \mathbb{C}^{N_t \times N_s}$, that maximizes the overall rate subject to the power constraint is given by matching to the set of eigenvectors corresponding to the $N_s$ largest eigenvalues of the channel covariance matrix, $\mathbf{S}[k] = {\mathbf{H}[k]}^{H}{\mathbf{H}[k]}$ \cite{telatar1999capacity}. \changet{In this section, we take advantage of the sparsity of the channel in \eqref{eq_ch_compact} to show that the channel covariance matrices in different subcarriers are approximately similar, hence they share approximately the same set of eigenvectors.}

For mmWave channel model \eqref{eq_ch_compact} when $N_t, N_r \to \infty$, the matrix ${\mathbf{S}[k]}$ can be further simplified as
%10
\begin{eqnarray} \nonumber
\label{eq_ch_asymptotic}
{\mathbf{S}[k]} &=&   \mathbf{A}_t \operatorname{diag}({\boldsymbol{\alpha}}^{*}[k]) \mathbf{A}_r^H \mathbf{A}_r \operatorname{diag}({\boldsymbol{\alpha}}[k]) \mathbf{A}_t^H \\
&\overset{(a)}{\approx}& \mathbf{A}_t \operatorname{diag}({\boldsymbol{\beta}}) \mathbf{A}_t^{H},
\end{eqnarray}
 where ${\boldsymbol{\beta}} = \left[|\alpha_{11}|^2,|\alpha_{12}|^2,\dots,|\alpha_{N_\text{c}N_\text{sc}}|^2\right]$. In \eqref{eq_ch_asymptotic}, the approximate equality $(a)$ is due to the fact that the diagonal elements of $\mathbf{A}_r^H \mathbf{A}_r$ are exactly equal to $1$ while the off-diagonal elements of that matrix are much smaller than $1$ with high probability \cite{el2013spatially}. Using a similar argument, we can show that $\mathbf{A}_t^H \mathbf{A}_t \approx \mathbf{I}$. Now, using this property and the structure of \eqref{eq_ch_asymptotic}, it can be seen that the columns of $\mathbf{A}_t$ are approximately the eigenvectors of $\mathbf{S}[k]$. 
 
Let $\tilde{\mathbf{A}}_t \in \mathbb{C}^{N_t \times N_s}$ denote the set of columns of $\mathbf{A}_t$ corresponding to $N_s$ largest $|\alpha_{c\ell}|^2$. It can be seen that the optimal fully-digital precoder in each subcarrier is now $\mathbf{V}_\text{opt}[k] = \tilde{\mathbf{A}}_t \boldsymbol{\Gamma}[k]$ where $\boldsymbol{\Gamma}[k]$ is the diagonal matrix of the allocated power to each data stream which can be obtained using water-filling approach. Since the elements of $\tilde{\mathbf{A}}_t$ typically satisfy the constant modulus norm constraint, those optimal fully-digital precoders can be realized by the hybrid precoding design in which $\mathbf{V}_\text{RF}= \tilde{\mathbf{A}}_t$  and $\mathbf{V}_D[k] =  \boldsymbol{\Gamma}[k]$. 

At the receiver side, with a similar justification, it is possible to show that the asymptotic optimal analog combiner is given by the set of columns of $\mathbf{A}_r$ corresponding to $N_s$ largest complex gains, $|\alpha_{c\ell}|^2$.

In summary, this section shows that in a mmWave SU-MIMO system with fixed number of scatterers in the environment and fixed number of data streams, if the number of antennas is sufficiently large so that $\mathbf{A}_t^H \mathbf{A}_t \approx \mathbf{I}$ and $\mathbf{A}_r^H \mathbf{A}_r \approx \mathbf{I}$, the hybrid beamforming with only $N_s$ RF chains can realize the optimal fully-digital beamforming.
% in which $\mathbf{A}_t^H \mathbf{A}_t \approx \mathbf{I}$ and $\mathbf{A}_r^H \mathbf{A}_r \approx \mathbf{I}$, 
%%%%%%%%%%%%%%%%%%%%%%%%%%%%%%%%%%%
%  Hybrid Beamforming Design
%%%%%%%%%%%%%%%%%%%%%%%%%%%%%%%%%%%
\section{Hybrid Beamforming Design for SU-MIMO}
\label{sec_design}
In the previous section, we show that the performance of the hybrid beamforming in frequency-selective channels can asymptotically approach the performance of the optimal fully-digital beamforming when the number of antennas at both transmitter and receiver is sufficiently large. Based on the asymptotic analysis, it is shown that the optimal analog beamforming design is to choose the columns of the analog precoder (combiner) from the transmit (receive) antenna array response vectors. However, as numerical results presented in Section~\ref{sec_sim} show, in scenarios with \changett{a} practical number of antennas in an environment with multiple scatters within each cluster, the performance of this asymptotic design actually has a sizable gap to the performance of fully-digital beamforming. Further, this asymptotic design requires fully-connected analog beamforming, and is not applicable to the partially-connected case. The goal of this section is to investigate how to design the hybrid beamformers for the scenarios with practical number of antennas and also with partially-connected structure.

\subsection{Transmitter Design}
In general, maximizing the spectral efficiency in \eqref{main_problem} requires a joint optimization over the transmit beamformers and receive beamformers. Joint optimization is, however, computationally complex. This paper uses the alternative strategy proposed in \cite{sohrabi2016hybrid,el2013spatially}, in which the transmitter is first designed assuming an ideal (fully-digital) receiver, then the receiver is designed given the already designed transmitter. This strategy leads to a decoupled transmitter and receiver design. 
Following this strategy, the hybrid beamforming design problem at the transmitter is
%11
\begin{subequations}
\label{main_problem_transmitter}
\begin{eqnarray}
&\displaystyle{\max_{\mathbf{V}_\text{RF},\{{\mathbf{V}_\text{D}[k]}\}_{k=1}^K}}  &\frac{1}{K} \displaystyle{\sum_{k=1}^{K}}
\tilde{R}[k] \\ 
\label{power_constraint_transmitter}
&\text{s.t.} &\operatorname{Tr}(\mathbf{V}_\text{RF} \mathbf{V}_\text{D}[k]
{\mathbf{V}_\text{D}[k]}^{H} \mathbf{V}_\text{RF}^{H}) \leq P\quad \\ 
\label{unity_const_matrix}
&&|\mathbf{V}_\text{RF}(i,j)|  = 1,  \forall (i,j) \in \mathcal{F}_t,
\end{eqnarray}
\end{subequations} 
where
%12
\begin{equation}
\label{rate_optimal_receiver}
\tilde{R}[k] = \log_2 \Bigl| \mathbf{I} + \frac{1}{\sigma^2} \mathbf{H}[k] \mathbf{V}_\text{RF} \mathbf{V}_\text{D}[k] {\mathbf{V}_\text{D}[k]}^{H} \mathbf{V}_\text{RF}^{H} {\mathbf{H}[k]}^{H}\Bigr|
\end{equation}
is the achievable rate of subcarrier $k$. This optimization problem is challenging since it is not convex even for single-carrier systems with flat-fading channels \cite{sohrabi2016hybrid}. To develop an algorithm for tackling this optimization problem, we take the following steps:
\begin{itemize}
\item First, for a fixed analog precoder, $\mathbf{V}_\text{RF}$, we derive the optimal closed-form solution for digital precoder of each subcarrier, $\mathbf{V}_\text{D}[k]$, that maximizes the overall spectral efficiency.
\item Second, by exploiting the structure of that optimal digital precoders, we show that the expression $\mathbf{V}_\text{D}[k]{\mathbf{V}_\text{D}[k]}^H$ in \eqref{rate_optimal_receiver} can be further simplified for large-scale antenna arrays.
\item Then, using that simplification and the Jensen's inequality, we derive an upper-bound for the objective function of \eqref{main_problem_transmitter}.
 \item  Finally, we devise an iterative algorithm to design the analog precoder such that it locally maximizes that upper-bound.
 \end{itemize}
%%%%%%%%%%%%%%%%%%%%%%%%%%%%%%%%%%%  
%		A- Digital Precoding Design
%%%%%%%%%%%%%%%%%%%%%%%%%%%%%%%%%%%
\subsubsection{Digital Precoding Design}
In this part, the optimal digital precoding design is presented given the analog precoder. When the analog precoder is fixed, the effective channel of subcarrier $k$ can be considered as $\mathbf{H}_\text{eff}[k]= \mathbf{H}[k]\mathbf{V}_\text{RF}$. 
Further, it can be seen that the constraints on digital precoders of different subcarriers in \eqref{main_problem_transmitter} are decoupled. Therefore, without loss of optimality, the following problem for designing the digital precoder of the subcarrier $k$ can be considered
%13
\begin{subequations}
\label{main_problem_digital_precoder} 
\begin{eqnarray}
&\displaystyle{\max_{ {\mathbf{V}_\text{D}[k]} } }  & \log_2 \Bigl| \mathbf{I} + \frac{1}{\sigma^2} \mathbf{H}_\text{eff}[k] \mathbf{V}_\text{D}[k] {\mathbf{V}_\text{D}[k]}^{H} {\mathbf{H}_\text{eff}[k]}^H \Bigr| \quad \\
&\text{s.t.} &\operatorname{Tr}(\mathbf{Q} \mathbf{V}_\text{D}[k]
{\mathbf{V}_\text{D}[k]}^{H} ) \leq P,
\label{power_constraint_dig_problem}
\end{eqnarray}
\end{subequations}
 where $\mathbf{Q} = \mathbf{V}_\text{RF}^{H}\mathbf{V}_\text{RF}$. The problem \eqref{main_problem_digital_precoder} has a closed-form water-filling solution as \cite{sohrabi2016hybrid}
%14
  \begin{equation}
  \label{digital_precoder_solution}
\mathbf{V}_\text{D}[k] = {\mathbf{Q}}^{-1/2} \mathbf{U}_e[k] \boldsymbol{\Gamma}_e[k],
\end{equation}
in which $\mathbf{U}_e[k]$ is the set of right singular vectors corresponding to the $N_s$ largest singular values of $\mathbf{H}_\text{eff}[k]\mathbf{Q}^{-1/2}$ and $\boldsymbol{\Gamma}_e[k]$ is the diagonal matrix of allocated powers to each symbol of the subcarrier $k$.

Now, we want to exploit the structure of the optimal digital precoder in \eqref{digital_precoder_solution} for large antenna arrays.
Toward this aim, we first present the following Lemma.
% LEMMA1
\begin{lemma}
\label{lem1}
Consider a hybrid beamforming transceiver with $N$ antenna elements, $N_\text{RF}$ RF chains and  the analog beamformer $\mathbf{F}_\text{RF}$. In fully-connected architecture, the analog beamformer satisfies $\mathbf{F}_\text{RF}^{H}\mathbf{F}_\text{RF} \approx N \mathbf{I}$ with high probability when $N\to \infty$, while in the partially-connected architecture the analog beamformer exactly satisfies $\mathbf{F}_\text{RF}^{H}\mathbf{F}_\text{RF} = \frac{N}{N_\text{RF}} \mathbf{I}, \forall N$.
\end{lemma}
\begin{IEEEproof}
In fully-connected architecture, the diagonal elements of $\mathbf{F}_\text{RF}^{H}\mathbf{F}_\text{RF}$ are exactly $N$ while the off-diagonal elements can be approximated as a summation of $N$ independent unit-norm complex numbers which implies that the norm of off-diagonal elements are much less than $N$ with high probability when $N \to \infty$. Therefore, the analog beamformer of fully-connected architecture typically satisfies $\mathbf{F}_\text{RF}^{H}\mathbf{F}_\text{RF} \approx N \mathbf{I}$.

In partially-connected architecture, the diagonal elements of $\mathbf{F}_\text{RF}^{H}\mathbf{F}_\text{RF}$ are $N/N_\text{RF}$, while the off-diagonal elements are exactly zero because of the block diagonal format of the analog beamformer (as in \eqref{partially_connected_structure}). So, the analog beamformer of partially-connected architecture exactly satisfies $\mathbf{F}_\text{RF}^{H}\mathbf{F}_\text{RF} = \frac{N}{N_\text{RF}} \mathbf{I}, \forall N$.  
\end{IEEEproof}

Now, using Lemma~\ref{lem1}, it is possible to simplify the expression of the optimal digital precoders in \eqref{digital_precoder_solution}. Based on Lemma~\ref{lem1}, the matrix $\mathbf{Q} = \mathbf{V}_\text{RF}^{H}\mathbf{V}_\text{RF}$ can always be approximated as proportional to the identity matrix,  $\mathbf{Q} \propto \mathbf{I}$. Moreover, for moderate and high SNR regime, we can adopt an equal power allocation for all streams in each subcarrier, ${\boldsymbol{\Gamma}}_e[k] \propto \mathbf{I}$, without significant performance degradation. Accordingly, the overall digital precoder can be approximated as 
 $\mathbf{V}_\text{D}[k]  \approx \gamma \mathbf{U}_e[k]$
 where
 $\gamma$ is a scalar parameter that guarantees that the power constraint in \eqref{power_constraint_dig_problem} is satisfied, i.e., for fully-connected structure $\gamma = \sqrt{P/(N_t N_\text{RF})} $ and for partially-connected structure $\gamma = \sqrt{P/N_t}$. 
%%%%%%%%%%%%%%%%%%%%%%%%%%%%%%%%%%%%%%%%%%%%%%%%%%%%%%%%%%%%%%%%%%%%%%
%	B. Analog Precoding Design
%%%%%%%%%%%%%%%%%%%%%%%%%%%%%%%%%%%%%%%%%%%%%%%%%%%%%%%%%%%%%%%%%%%%%%
\subsubsection{Analog Precoding Design}
\label{RF_precoder_design} We now present an algorithm for designing the analog precoder assuming that the digital precoder in each subcarrier is given as $\mathbf{V}_\text{D}[k]  \approx \gamma \mathbf{U}_e[k]$. It can be seen that for such digital precoder we have 
%15
  \begin{equation}
  \label{VD_structure}
  \mathbf{V}_\text{D}[k] {\mathbf{V}_\text{D}[k]}^H \approx \gamma^2 \mathbf{U}[k]  \tilde{\mathbf{I}}_{N_\text{RF}}
 \mathbf{U}[k]^H,
\end{equation}
where $\tilde{\mathbf{I}}_{N_\text{RF}} = \begin{bmatrix}
\mathbf{I}_{N_s} & \mathbf{0}\\
\mathbf{0} & \mathbf{0}
 \end{bmatrix}$ and $\mathbf{U}[k]\in \mathbb{C}^{N_\text{RF}\times N_\text{RF} }$ is a unitary matrix. Therefore, the achievable rate of subcarrier $k$ in \eqref{rate_optimal_receiver} can be upper-bounded as
%16
\begin{eqnarray}
 \nonumber
&\tilde{R}[k] &= \log_2 \Bigl| \mathbf{I} + \frac{\gamma^2}{\sigma^2} \mathbf{U}[k]^H \mathbf{V}_\text{RF}^{H} {\mathbf{H}[k]}^{H} \mathbf{H}[k] \mathbf{V}_\text{RF} \mathbf{U}[k] \tilde{\mathbf{I}}_{N_\text{RF}}\Bigr|\\ \nonumber
&&\overset{(a)}{\leq} \log_2 \Bigl| \mathbf{I} + \frac{\gamma^2}{\sigma^2} \mathbf{U}[k]^H \mathbf{V}_\text{RF}^{H} {\mathbf{H}[k]}^{H} \mathbf{H}[k] \mathbf{V}_\text{RF} \mathbf{U}[k] {\mathbf{I}}_{N_\text{RF}}\Bigr|\\ 
 && \overset{(b)}{=} \log_2 \Bigl| \mathbf{I}  + \frac{\gamma^2}{\sigma^2} \mathbf{V}_\text{RF}^{H} {\mathbf{H}[k]}^{H} \mathbf{H}[k] \mathbf{V}_\text{RF} \Bigr|,
 \label{upper_per_sub}
\end{eqnarray}
where $(a)$ is satisfied with equality if $N_\text{RF} = N_s$ and $(b)$ is due to the properties of the unitary matrices. Finally, we derive the upper-bound for the overall spectral efficiency in the objective of \eqref{main_problem_transmitter} using Jensen's inequality as
%17
\begin{eqnarray}
 \nonumber
&\displaystyle{ \frac{1}{K}} \displaystyle{\sum_{k=1}^{K}}\tilde{R}[k] & \overset{(c)}{\leq} \frac{1}{K} \displaystyle{\sum_{k=1}^{K}}\log_2 \Bigl| \mathbf{I}  + \frac{\gamma^2}{\sigma^2} \mathbf{V}_\text{RF}^{H} {\mathbf{H}[k]}^{H} \mathbf{H}[k] \mathbf{V}_\text{RF} \Bigr|\\
&& \overset{(d)}{\leq} \log_2 \Bigl| \mathbf{I}  + \frac{\gamma^2}{\sigma^2} \mathbf{V}_\text{RF}^{H}  \mathbf{F}_1 \mathbf{V}_\text{RF} \Bigr|,
\label{eq_up}
\end{eqnarray}
where 
%18
\begin{equation}
\mathbf{F}_1 = \frac{1}{K} \displaystyle{\sum_{k=1}^{K}}\left({\mathbf{H}[k]}^{H} \mathbf{H}[k]\right),
\end{equation}
is the average of the covariance matrices of frequency domain channels, $(c)$ follows \eqref{upper_per_sub} and $(d)$ is based on Jensen's inequality; i.e., for a concave function $f(\cdot)$, if $\sum_i \alpha_i = 1$, then $\sum_i \alpha_i f(\mathbf{X}_i) \leq f(\sum_i \alpha_i \mathbf{X}_i)$.

In this paper, we propose to design the analog precoder such that it maximizes the upper-bound of the overall spectral efficiency in \eqref{eq_up}, yielding
%19
\begin{subequations}
\label{main_problem_transmitter_RF}
\begin{eqnarray}
\label{objective_transmitter_RF}
 &\displaystyle{\max_{\mathbf{V}_\text{RF}}}  &\log_2 \Bigl| \mathbf{I}  + \frac{\gamma^2}{\sigma^2} \mathbf{V}_\text{RF}^{H} \mathbf{F}_1 \mathbf{V}_\text{RF} \Bigr|\\
\label{unity_const_matrix_RF}
&\text{s.t.} &|\mathbf{V}_\text{RF}(i,j)|  = 1,  \forall (i,j)\in \mathcal{F}_t.
\end{eqnarray}
\end{subequations}

Interestingly, the problem \eqref{main_problem_transmitter_RF} is now in the format of analog precoder design problem for single-carrier systems with flat-fading channels which we have considered in \cite{sohrabi2016hybrid}. In \cite{sohrabi2016hybrid}, we show that the analog precoder of a single-carrier system should be designed to maximize ${\log_2 \Bigl| \mathbf{I}  + \frac{\gamma^2}{\sigma^2} \mathbf{V}_\text{RF}^{H} ( {\mathbf{H}}^{H} \mathbf{H}) \mathbf{V}_\text{RF} \Bigr|}$.
Now, the optimization problem in \eqref{main_problem_transmitter_RF} suggests that we can replace the covariance matrix of the channel, ${\mathbf{H}}^{H} \mathbf{H}$, in flat-fading scenario by its average over all subcarriers, i.e., $\mathbf{F}_1 = \frac{1}{K} {\sum_{k=1}^{K}}\left({\mathbf{H}[k]}^{H} \mathbf{H}[k]\right)$, and then use the algorithm in \cite{sohrabi2016hybrid} for designing the analog precoder in OFDM-based systems with frequency-selective channels. The rest of this section provides a brief explanation of the iterative algorithm in \cite{sohrabi2016hybrid}.

It can be seen that all the constraints in problem \eqref{main_problem_transmitter_RF} are decoupled. This enables us to develop an iterative coordinate descent algorithm over the elements of analog precoder, $\mathbf{V}_\text{RF}$, to find a locally-optimal solution of the problem \eqref{main_problem_transmitter_RF}. Mathematically, it is shown in \cite{sohrabi2016hybrid} that the contribution of each element of analog precoder, $\mathbf{V}_\text{RF}(i,j)$, to the objective of \eqref{main_problem_transmitter_RF} can be extracted as
 %20
\begin{equation}
\label{isolation}
\log_2 \bigl| \mathbf{C}_j \bigr| + \log_2 \left( 2\operatorname{Re}\bigl\{ {{\mathbf{V}_\text{RF}^{*}}(i,j)} \eta_{ij} \bigr\} + \zeta_{ij} +1\right),
\end{equation}
where $ \mathbf{C}_j = \mathbf{I} + \frac{\gamma^2}{\sigma^2} { 
({{\bar{\mathbf{V}}}_\text{RF}^j})^H   
\mathbf{F}_1 {\bar{\mathbf{V}}_\text{RF}^{j}}}$,
and 
${\bar{\mathbf{V}}_\text{RF}^{j}}$
  is the sub-matrix of $\mathbf{V}_\text{RF}$ with $j^\text{th}$ column removed,
  and
$\eta_{ij} = {{\sum_{\ell\not=i}} } \mathbf{G}_j(i,\ell) {\mathbf{V}_\text{RF}(\ell,j)}$, 
and
$\zeta_{ij} =  \mathbf{G}_j(i,i) +2\operatorname{Re} \left\{ {\sum_{m\not=i, n\not=i}}{{{\mathbf{V}_\text{RF}^{*}}(m,j)}}\mathbf{G}_j(m,n) {\mathbf{V}_\text{RF}(n,j)}  \right\} $, 
and
$\mathbf{G}_j =  \frac{\gamma^2}{\sigma^2} \mathbf{F}_1 - \frac{\gamma^4}{\sigma^4} \mathbf{F}_1   {\bar{\mathbf{V}}_\text{RF}^{j}}  \mathbf{C}_j^{-1} ({\bar{\mathbf{V}}_\text{RF}^{j}})^H   \mathbf{F}_1 $. 
 Now, because all parameters $\mathbf{C}_j$, $\zeta_{ij}$ and $\eta_{ij}$ are independent of the element $\mathbf{V}_\text{RF}(i,j)$, the optimal value for this element (when all other elements of $\mathbf{V}_\text{RF}$ are fixed) is given as
%21
\begin{align}
\label{sequential_update_PTP}
\mathbf{V}_\text{RF}(i,j) =   
\begin{cases} \frac{\eta_{ij}}{|\eta_{ij}|}, & \mbox{$\forall(i,j)\in\mathcal{F}_t$ s.t. $\eta_{ij}\not=0$},\\ 
1, & \mbox{$\forall(i,j)\in\mathcal{F}_t$ s.t. $\eta_{ij}=0$},\\ 
0, &\mbox{$\forall(i,j)\notin\mathcal{F}_t$}.\end{cases}
\end{align}

The final algorithm consists of starting with initial feasible analog precoder which satisfies \eqref{unity_const_matrix_RF}, then sequentially updating each $\mathbf{V}_\text{RF}(i,j)$ based on \eqref{sequential_update_PTP}. The convergence of this algorithm to the local optimal solution of \eqref{main_problem_transmitter_RF} is guaranteed since in each step of algorithm the objective function increases.
%%%%%%%%%%%%%%%%%%%%%%%%%%%%%%%%%%%
%		C- Receiver Design
%%%%%%%%%%%%%%%%%%%%%%%%%%%%%%%%%%%
\subsection{Receiver Design}
\label{V}
We now consider the hybrid combining design when the transmit beamforming matrix is already fixed. For a fixed analog combiner, the optimal digital combiner of each subcarrier is known to be the MMSE solution as
%22
\begin{equation}
\label{digital_com_solution}
\mathbf{W}_\text{D}[k] = {\mathbf{J}[k]}^{-1} \mathbf{W}_\text{RF}^H \mathbf{H}[k] \mathbf{V}_\text{t}[k],
\end{equation}
in which $\mathbf{J}[k] = \mathbf{W}_\text{RF}^H \mathbf{H}[k] \mathbf{V}_\text{t}[k] \mathbf{V}_\text{t}[k]^H \mathbf{H}[k]^H \mathbf{W}_\text{RF} + \sigma^2 \mathbf{W}_\text{RF}^H \mathbf{W}_\text{RF} $.
According to Lemma~\ref{lem1}, the analog combiner satisfies $\mathbf{W}_\text{RF}^H \mathbf{W}_\text{RF}. \propto \mathbf{I}$, for both fully-connected and partially-connected structures. This leads to the conclusion that the effective noise after the analog combiner approximately remains white. Now, using the property of MMSE digital combiner under white background noise, the analog combiner design problem can be written as \cite{sohrabi2016hybrid}
%23
\begin{subequations}
\label{problem_receiver_TF_PTP} 
\begin{eqnarray}
&\displaystyle{\max_{\mathbf{W}_\text{RF}}} \hspace{-8pt}&\frac{1}{K} \sum_{k=1}^K \log \Bigl| \mathbf{I}  +  (\sigma^2 \mathbf{W}_\text{RF}^H \mathbf{W}_\text{RF})^{-1}\mathbf{W}_\text{RF}^H \tilde{\mathbf{F}}[k] \mathbf{W}_\text{RF}\Bigr|  ~~~~~~~\\
&\text{s.t.} \hspace{-8pt}&|\mathbf{W}_\text{RF}(i,j)|^2 = 1, \forall (i,j) \in \mathcal{F}_r,
\end{eqnarray}
\end{subequations}
where $\tilde{\mathbf{F}}[k] = \mathbf{H}[k] \mathbf{V}_\text{t}[k] \mathbf{V}_\text{t}[k]^{H} {\mathbf{H}[k]}^{H}$.
 By using Lemma~\ref{lem1}, $\mathbf{W}_\text{RF}^H \mathbf{W}_\text{RF} \propto \mathbf{I}$, and applying the Jensen's inequality, we can consider maximizing the upper-bound of \eqref{problem_receiver_TF_PTP} for designing $\mathbf{W}_\text{RF}$ as
%24
\begin{subequations}
\begin{eqnarray} 
&\displaystyle{\max_{\mathbf{W}_\text{RF}}}  & \log_2 \Bigl| \mathbf{I}  + \frac{1}{\sigma^2 \tau} \mathbf{W}_\text{RF}^H {\mathbf{F}}_2 \mathbf{W}_\text{RF}\Bigr|  \\
&\text{s.t.} &|\mathbf{W}_\text{RF}(i,j)|^2 = 1, \forall(i,j) \in \mathcal{F}_r,
\end{eqnarray}
\end{subequations}
in which $\mathbf{F}_2 = \frac{1}{K}\sum_{k=1}^K \tilde{\mathbf{F}}[k]$, $\tau = N_r$ and $\tau = N_r/N_\text{RF}$ for fully-connected structure and partially-connected structure, respectively. It can be seen that this problem is in the same format as analog precoder design problem in \eqref{main_problem_transmitter_RF}. Therefore, the analog combiner $\mathbf{W}_\text{RF}$ can be designed using the proposed algorithm in Section \ref{RF_precoder_design}. \changet{We note that the receiver design is based on the already designed transmitter. This implies that
either a dedicated phase is required to feed forward the designed transmit beamformers to the receiver, or it is required that the receiver first solves the transmitter design problem and then solves the receiver design problem. In the former case, the extra radio communication resource is needed while in the latter case extra computation resource is needed at the receiver.}

%%%%%%%%%%%%%%%%%%%%%%%%%%%%%%%%%%%
%Algorithm 1
%%%%%%%%%%%%%%%%%%%%%%%%%%%%%%%%%%%
\begin{algorithm}[t]
\caption{Design of Hybrid Beamformers for SU-MIMO systems}
\label{Alg:hehe}
%\textbf{Require:} $\sigma^2$, $P$, $K$,$\mathbf{H}[k]$
\begin{algorithmic}[1]
	\State Find $\mathbf{V}_\text{RF}$ by solving the problem in \eqref{main_problem_transmitter_RF}
	using Algorithm~1 in \cite{sohrabi2016hybrid}.
	\State Calculate $\mathbf{V}_\text{D}[k] = (\mathbf{V}_\text{RF}^H \mathbf{V}_\text{RF})^{-1/2} \mathbf{U}_e[k] \boldsymbol{\Gamma}_e[k]$ where $\mathbf{U}_e[k]$ and $\boldsymbol{\Gamma}_e[k]$ are defined as following \eqref{digital_precoder_solution}.
	\State Find $\mathbf{W}_\text{RF}$ by solving the problem in \eqref{problem_receiver_TF_PTP}  using Algorithm~1  in \cite{sohrabi2016hybrid}.
	\State Calculate $\mathbf{W}_\text{D}[k] = \mathbf{J}[k]^{-1} \mathbf{W}_\text{RF}^H \mathbf{H}[k] \mathbf{V}_\text{RF}\mathbf{V}_\text{D}[k]$ where $\mathbf{J}[k]$ is defined as following \eqref{digital_com_solution}.
\end{algorithmic}
\end{algorithm}

The summary of the overall proposed algorithm for the hybrid beamforming design to maximize the overall spectral efficiency in an OFDM-based large-scale SU-MIMO system is given in 
Algorithm~\ref{Alg:hehe}. Assuming that the number of antennas at both ends are in the same range, i.e., $N_r = O(N_t)$, it can be shown the computational complexity of the overall algorithm is $O(KN_t^3)$.
\begin{A_remark}
\normalfont
So far, we assume that the infinite-resolution phase shifters are available at the transceivers. However, in practice the components required for such an accurate phase control may be costly \cite{krieger2013dense}. So an interesting question is how to design the analog beamformers if only low-resolution phase shifters are available; i.e., $\mathbf{V}_\text{RF}(i,j) \in \mathcal{G}$ and $\mathbf{W}_\text{RF}(i,j) \in \mathcal{G}$ where $\mathcal{G} = \{ 1,\omega, \omega^2, \dots \omega^{2^b-1} \}$, $\omega = e^{j\frac{2\pi}{2^b}}$ and $b$ is the number of bits in the resolution of phase shifters. We have already addressed this problem in \cite{sohrabi2016hybrid} for single-carrier system by quantizing the solution of the analog precoder element in \eqref{sequential_update_PTP} in each iteration to the closest point in $\mathcal{G}$. Since the preceding treatment essentially transforms the multi-carrier analog beamformer design problem to the format of analog precoder design problem for the single-carrier scenario, we can use the same technique. 
\end{A_remark}
%%%%%%%%%%%%%%%%%%%%%%%%%%%%%%%%%%%
%  Hybrid Precoding Design for MU-MISO
%%%%%%%%%%%%%%%%%%%%%%%%%%%%%%%%%%%
\section{Hybrid Precoding Design for MU-MISO}
\label{sec:miso}
We now consider the hybrid precoding design for a OFDM-based MU-MISO system in which a base station with $N_t$ antennas and $N_\text{RF}$ RF chains serves $N_U$ non-cooperative single-antenna users. It is shown in \cite{kong2015wideband} that hybrid beamforming with limited number of RF chains can approach the performance of fully-digital beamforming only if the same set of users are scheduled to be served at the entire band for each time slot. This is because when the same users are served in the entire band, the channels of different subcarriers are highly correlated due to the sparse nature of mmWave channels. Accordingly, it is possible to design a common analog beamformer which is appropriate for all the channels. This is more spectrally efficient than the alternative of multiplexing users across the frequencies. For this reason, this section focuses on a MU-MISO design in which the same users are served over all subcarriers.

For such a system, the transmitted signal at subcarrier $k$ is 
%25
\begin{equation}
\mathbf{x}[k] = \sum_{i=1}^{N_U}\mathbf{V}_\text{RF} \mathbf{v}_{\text{D}_i}[k] s_i[k],
\end{equation}
where $\mathbf{V}_\text{RF}\in \mathbb{C}^{N_t\times N_\text{RF}}$ is the analog precoder, $\mathbf{v}_{\text{D}_i}[k] \in \mathbb{C}^{N_\text{RF}\times 1}$ is the digital precoder for user $i$ at subcarrier $k$ and  $s_i[k]\in \mathbb{C}$ is the intended data symbol for user $i$ at subcarrier $k$. Then, the user $n$ receives $y_n[k] = \mathbf{h}_n^H[k] \mathbf{x}[k] + z_n[k]$, where $z_n[k]$ is the additive white Gaussian noise. The rate expression in $k^\text{th}$ subcarrier for user $n$ in \eqref{rate} can be expressed as
%26
\begin{equation}
R_n[k] = \log_2 \left( 1 + \frac{\big|\mathbf{h}_n^H[k] \mathbf{V}_\text{RF} \mathbf{v}_{\text{D}_n}[k]\big|^2 } {\sigma^2 + \sum_{i \not= n}  \big|\mathbf{h}_n^H[k] \mathbf{V}_\text{RF} \mathbf{v}_{\text{D}_i}[k]\big|^2 } \right),
\end{equation}
where $\mathbf{h}_n^H[k]$ is the channel of $k^{th}$ subcarrier  from the BS to the $n^\text{th}$ user. Now, the hybrid precoding design problem for maximizing the weighted sum rate is
%27
\begin{subequations}
\label{main_problem_transmitterMISO}
\begin{eqnarray}
&\displaystyle{\max_{\mathbf{V}_\text{RF},\{{\mathbf{V}_\text{D}[k]}\}_{k=1}^K}}  &\frac{1}{K} \displaystyle{\sum_{k=1}^{K}} ~ \displaystyle{\sum_{n=1}^{N_U}}
\beta_n {R}_n[k] \\ 
\label{27_power}
&\text{s.t.} &\operatorname{Tr}(\mathbf{V}_\text{RF} \mathbf{V}_\text{D}[k]
{\mathbf{V}_\text{D}[k]}^{H} \mathbf{V}_\text{RF}^{H}) \leq P\quad \\ 
&&|\mathbf{V}_\text{RF}(i,j)|  = 1,  \forall (i,j) \in \mathcal{F}_t,
\end{eqnarray}
\end{subequations} 
in which $\mathbf{V}_D[k] = \left[\mathbf{v}_{\text{D}_1}[k],\dots,\mathbf{v}_{\text{D}_{N_U}}[k] \right]$ is the overall digital precoder at $k^{th}$ frequency tone and the weight $\beta_n$ represents the priority of $n^\text{th}$ user; i.e., larger ${\beta_n}$ implies greater priority for $n^\text{th}$ user.

  The problem of weighted sum rate maximization in \eqref{main_problem_transmitterMISO} differs from the spectral efficiency maximization for the SU-MIMO systems in two respects. First, the users in MU-MISO scenario are not collocated  which results in an inter-user interference term in the rate expression. Second, the different data streams corresponding to different users in a MU-MISO system may have different priority weights, while all the data streams in a SU-MIMO system always have the same priority weights. 
  
These two differences make the analog precoding design much more complicated. In order to tackle the problem \eqref{main_problem_transmitterMISO}, this paper proposes the following simple design strategy:
 \begin{itemize}
 \item First design the analog precoder assuming that the users are cooperative and they have equal priority weights. \changet{Loosely speaking, this means that the common analog beamformer is designed to improve the direct channel of the all users while neglecting the effect of the inter-user interference and the different priority weights.}
 
 \item  Second, design the digital beamformers for the effective channel using one of the conventional fully-digital beamforming schemes such as the WMMSE approach in \cite{shi2011iteratively} which can manage both the inter-user interference effect and the different priority weights. 
  \end{itemize} 
  
In particular, we first consider designing the analog precoder by maximizing $\log_2 \Bigl| \mathbf{I}  + \frac{\gamma^2}{\sigma^2} \mathbf{V}_\text{RF}^{H} \mathbf{F}_1 \mathbf{V}_\text{RF} \Bigr|$ where $\mathbf{F}_1 = \frac{1}{K} {\sum_{k=1}^{K}}\left({\mathbf{H}[k]}^{H} \mathbf{H}[k]\right)$ and $\mathbf{H}[k] = \bigl[\mathbf{h}_1[k],\dots,\mathbf{h}_{N_U}[k]\bigr]^H$ is the collection of channel vectors of all users at subcarrier $k$. It is clear that the analog precoder now can be designed using the proposed algorithm in Section~\ref{sec_design}. After designing the analog precoder, we seek to design the digital precoders using the iterative WMMSE approach which is summarized as follows:
%%%%%% Algorithm Digital precoder in MU-MISO
\begin{enumerate}
 \item Let  $\mathbf{g}_n^H[k] = \mathbf{h}_n^H[k] \mathbf{V}_\text{RF}$ denote the effective channel. Initialize all digital precoders, $\mathbf{v}_{\text{D}_n}[k], \forall n,k$, such that the power constraints \eqref{27_power} are satisfied.
  \item Calculate the receiver combining filter for each user at each subcarrier as $w_n[k] = \frac{ \mathbf{g}_n^H[k] \mathbf{v}_{\text{D}_n}[k] } {\sigma^2 + \sum_{i}  \left|\mathbf{g}_n^H[k]  \mathbf{v}_{\text{D}_i}[k]\right|^2 }$.
  \item Calculate the mean square error (MSE) as
  \begin{multline*}
	e_n[k] = \sum_i \big|w_n[k] \mathbf{g}_n^H[k] \mathbf{v}_{\text{D}_i}[k]\big|^2 \\
	-2\operatorname{Re}\{w_n^*[k] \mathbf{g}_n^H[k] \mathbf{v}_{\text{D}_i}  \}+ |w_n[k]|^2\sigma^2 + 1,
	\end{multline*}
	and then calculate $t_n[k] = \frac{1}{e_n[k]}$.
\item Design the digital precoder as 
  \begin{equation} \nonumber
	\mathbf{v}_{\text{D}_n}[k] = \beta_n t_n[k]w_n[k] \left( \mathbf{J}_n[k] \right)^{-1} \mathbf{g}_n[k],
	\end{equation}
$\mathbf{J}_n[k] = {\sum_i} \beta_i t_i[k]\left|w_i[k]\right|^2 \mathbf{g}_i[k] \mathbf{g}_i^H[k]  + \lambda[k] \mathbf{V}_\text{RF}^H \mathbf{V}_\text{RF}$ where  $\lambda[k]$ is the Lagrangian multiplier for subcarrier $k$ which can be optimized based on Karush-Kuhn-Tucker (KKT) conditions \cite{shi2011iteratively}.
\item Repeat the steps from 2 to 5 until the convergence.
\end{enumerate}

Now, assuming that $\min\{K,N_t\} \gg N_\text{RF}$, it can be shown that the dominant term in computational cost is $O(KN_t^2)$ which corresponds to the calculation of the average of the covariance matrices of frequency domain channels,  $\frac{1}{K} {\sum_{k=1}^{K}}\left({\mathbf{H}[k]}^{H} \mathbf{H}[k]\right)$. As a result, the overall computational complexity of the proposed hybrid beamforming algorithm is $O(KN_t^2)$.

 We note that the recent work in \cite{kong2015wideband} also considers hybrid precoding design for weighted sum rate maximization in a OFDM MU-MISO system. The general algorithm in \cite{kong2015wideband} uses the WMMSE technique  to design the analog precoder as well as the digital precoders. The algorithm of \cite{kong2015wideband} can only be applied to the setting where the total power constraint is considered, $\sum_k \operatorname{Tr}(\mathbf{V}_\text{RF} \mathbf{V}_\text{D}[k]
{\mathbf{V}_\text{D}[k]}^{H} \mathbf{V}_\text{RF}^{H}) \leq KP$. However, in practical systems it is desirable that the designed precoders satisfy per subcarrier power constraint \eqref{27_power}. The reason that it is difficult to generalize the algorithm of \cite{kong2015wideband} for the per subcarrier power constraint case is that in the
analog precoder design step within WMMSE approach, multiple Lagrangian multipliers would arise, and the resulting optimization problem would be computationally difficult to solve. \changet{We also note that although the focus of this paper is on hybrid beamforming design for power spectral density constraint, the proposed algorithm can be easily modified for the total power constraint scenario since we use WMMSE approach only to design digital precoders. It can be shown that this approach is more computationally efficient as compared to the algorithm in \cite{kong2015wideband} since the analog beamformer in \cite{kong2015wideband} needs to be updated in each iteration of WMMSE method and the expression of the analog beamformer involves a large dimension matrix inversion. Further, since the design of digital and analog precoders are decoupled in the proposed algorithm, when the analog precoder is fixed, other simpler linear beamforming approaches such as zero-forcing and maximum ratio transmission can be employed as the digital precoders to reduce the design complexity.}

%%%%%%%%%%%%%%%%%%%%%%%%%%%%%%%%%%%%%%
%%%%Simulations
%%%%%%%%%%%%%%%%%%%%%%%%%%%%%%%%%%%%%%
\section{Simulations}
\label{sec_sim}
In this section, we present numerical simulation results for both the asymptotic hybrid beamforming design presented in Section~\ref{sec_asy} and the proposed algorithms 
for SU-MIMO systems and MU-MISO systems presented in Section~\ref{sec_design} and Section~\ref{sec:miso}.
In the simulations, we consider the uniform linear array antenna configuration with half-wavelength antenna spacing. Further, unless otherwise mentioned, we consider an environment with $5$ clusters and $10$ scatterers per cluster \cite{akdeniz2014millimeter} in which the angles of arrival (departure) are generated according to Laplacian distribution with random mean cluster angels $\bar{\phi}_{c}^r \in [0,2\pi)$ ($\bar{\phi}_{c}^t \in [0,2\pi)$) and angular spreads of $10$ degrees within each cluster. In simulations for SU-MIMO systems, the average spectral efficiency is plotted versus the number of antenna elements or the signal-to-noise-ratio per subcarrier ($\operatorname{SNR}=\frac{P}{\sigma^2}$) over $100$ channel realizations as a performance metric.

%%%%%%%%%%%%%%%%%%%%%%%%%%%%%%%%%%%%
%% Fig. 3
%%%%%%%%%%%%%%%%%%%%%%%%%%%%%%%%%%%%
\begin{figure}
    \centering
    \begin{subfigure}[b]{.51\textwidth}
        \includegraphics[width=\textwidth]{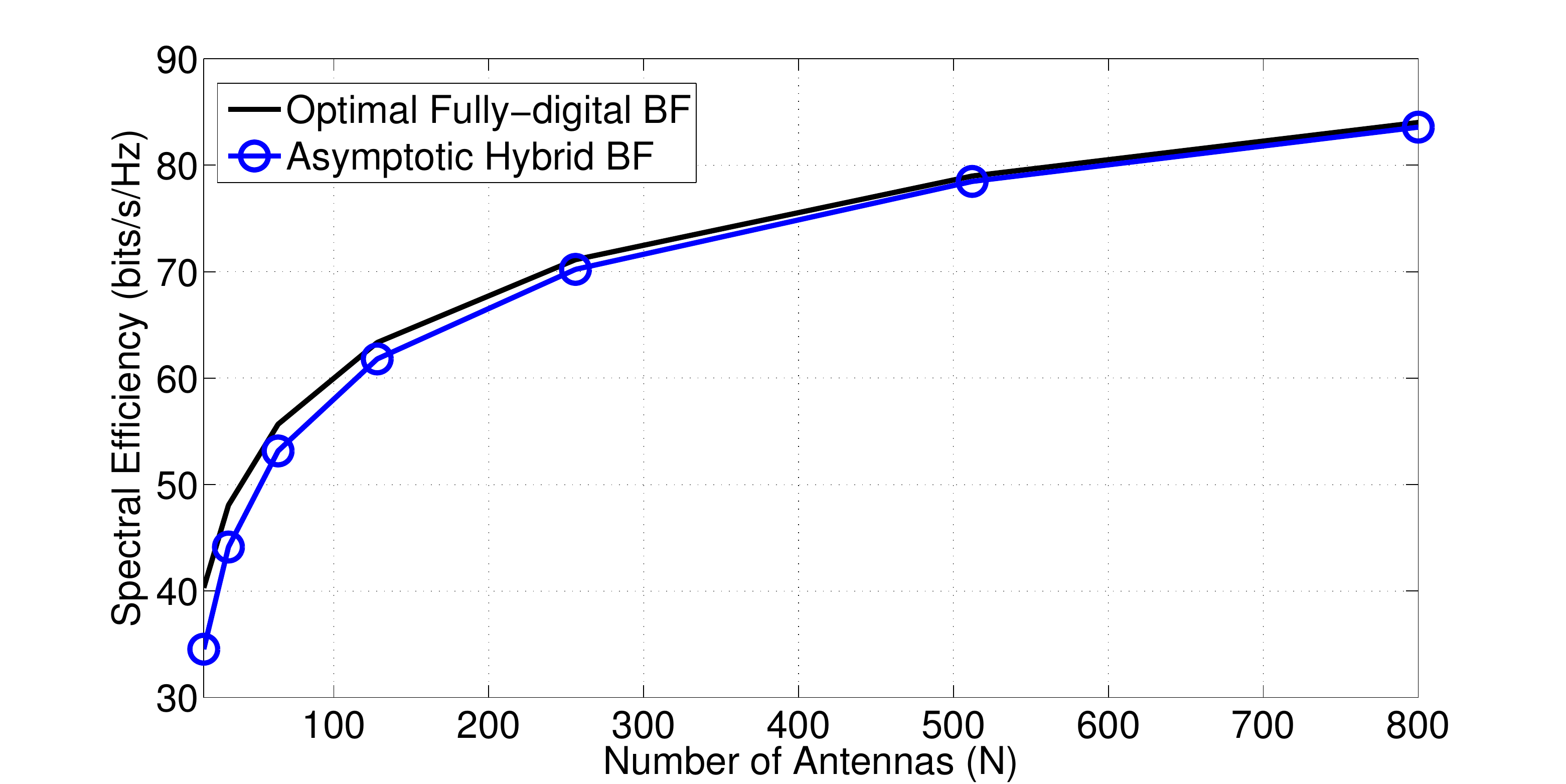}
        \caption{mmWave channel model with $(N_\text{c},N_\text{sc}) = (15,1)$}
        \label{fig_single_sc}
    \end{subfigure}
    ~ 
    \begin{subfigure}[b]{0.51\textwidth}
        \includegraphics[width=\textwidth]{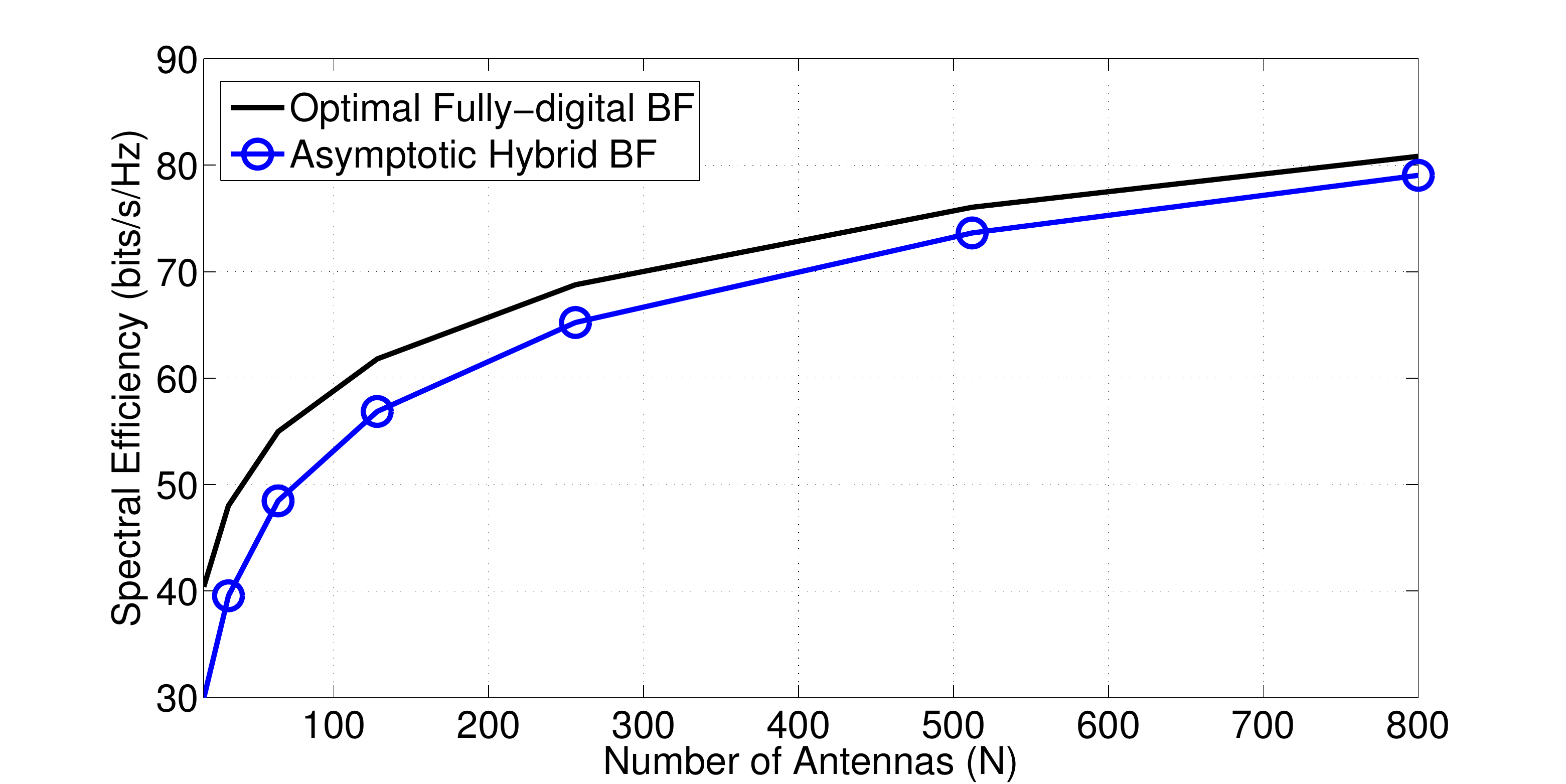}
        \caption{mmWave channel model with $(N_\text{c},N_\text{sc}) = (5,10)$}
        \label{fig_multi_sc}
    \end{subfigure}
    \caption{Comparison between the achievable rates of asymptotic design and the optimal fully-digital beamforming for an $N\times N$ MIMO system with $N_s = N_\text{RF} =4$, $K=32$ and \changet{$\operatorname{SNR} = 20$dB}.}\label{fig:system_models}
\end{figure}

%%%%%%%%%%%%%%%%%%%%%%%%%%%%%%%%%%%%%
%				Asymptotic Hybrid Beamforming Design Analysis for SU-MIMO systems
%%%%%%%%%%%%%%%%%%%%%%%%%%%%%%%%%%%%%
\subsection{Asymptotic Hybrid Beamforming Design Analysis for SU-MIMO systems}
First, we numerically investigate the performance of asymptotic hybrid beamforming design of Section~\ref{sec_asy} for large antenna arrays in different propagation environments. Here, we consider an $N\times N$ MIMO system with \changet{$\operatorname{SNR} = 20$dB} and $K=32$ subcarriers in which a transmitter with $4$ RF chains sends $N_s = 4$ data streams per subcarrier to a receiver with $4$ RF chains. To model the propagation environment, two scenarios are considered, single-scatterer per cluster and multi-scatterers per cluster. In 
%Fig.~\ref{fig_single_sc} and Fig.~\ref{fig_multi_sc},
Fig.~3(a) and Fig.~3(b), 
the performance of the asymptotic hybrid beamforming design is compared to that of the optimal fully-digital beamforming by sweeping the number of antenna elements, $N$, in environments with $(N_\text{c},N_\text{sc}) = (15,1)$ and  $(N_\text{c},N_\text{sc}) = (5,10)$, respectively. It can be seen in both cases that the achievable rate of asymptotic design converges to that of the fully-digital beamforming for sufficiently large number of antennas. However, this convergence is faster in single-scatterer per cluster scenario. This is because in single-scatterer per cluster case the angle of arrivals (departures) are independent and hence the asymptotic orthogonality of the columns of receive (transmit) antenna array response matrix, $\mathbf{A}_r$ ($\mathbf{A}_t$), is valid for smaller values of $N$.
%%%%%%%%%%%%%%%%%%%%%%%%%%%%%%%%%%%%%
%				Hybrid Beamforming Analysis in MIMO systems
%%%%%%%%%%%%%%%%%%%%%%%%%%%%%%%%%%%%%
\subsection{Hybrid Beamforming Analysis in SU-MIMO systems}
Next, we numerically evaluate the performance of the hybrid beamforming design presented in Section~\ref{sec_design} for OFDM-based SU-MIMO systems in two different system settings:

 \subsubsection{SU-MIMO system with hybrid beamforming architecture at both transceiver sides}
\begin{figure}[t]
	\centering
	\includegraphics[width=0.51\textwidth]{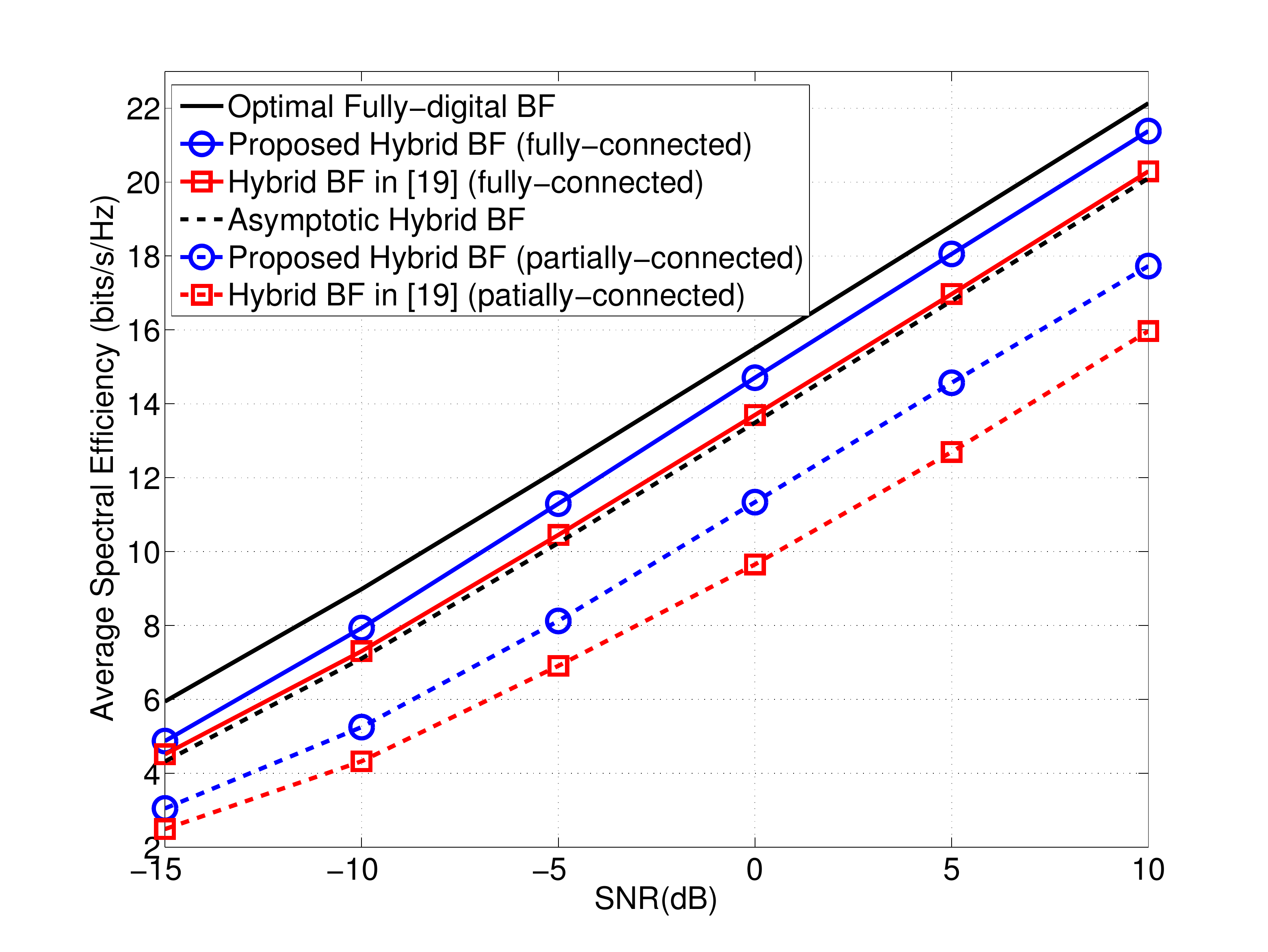}
	\centering
	\caption{Spectral efficiencies versus SNR of different methods for a $64\times 32$ OFDM-based SU-MIMO system in which hybrid architecture with infinite resolution phase shifters is employed at both transceiver sides and $N_s=2$, $N_\text{RF} =4$ and $K=64$.
	}
	\label{Fig_VsAlt}
\end{figure}

First, consider a $64 \times 32$ OFDM-based MIMO system with hybrid beamforming structure at both the transmitter and the receiver where $N_s = 2$, $N_\text{RF} = 4$ and $K=64$. Fig.~\ref{Fig_VsAlt} shows that the proposed algorithm for both fully-connected and partially-connected structures achieves a higher spectral efficacy as compared to the hybrid beamforming design in \cite{yu2016alternating} which seeks to minimize the distance of the optimal fully-digital beamformers and the overall hybrid beamformers instead of tackling the original problem of spectral efficiency maximization directly.  Fig.~\ref{Fig_VsAlt} also shows that the proposed fully-connected hybrid beamforming design with only $4$ transceiver RF chains can approach the performance of fully-digital beamforming which is achieved by the optimal fully-digital beamforming utilizing $64$ and $32$ RF chains at the transmitter and the receiver, respectively. Finally, Fig.~\ref{Fig_VsAlt} indicates that for the scenarios in which the number of antenna is not extremely large the proposed algorithm in Section~\ref{sec_design} can achieve a better performance; about $2$dB gain, in comparison with the  asymptotic design of  Section~\ref{sec_asy}.

%%%%%%%%%%%%%%%%%%%%%%%%%%%%%%%%%%%%
%% Fig. 4
%%%%%%%%%%%%%%%%%%%%%%%%%%%%%%%%%%%%

 \subsubsection{SU-MIMO system with hybrid beamforming architecture only at the transmitter side}
\begin{figure}[t]
	\centering
	\includegraphics[width=0.51\textwidth]{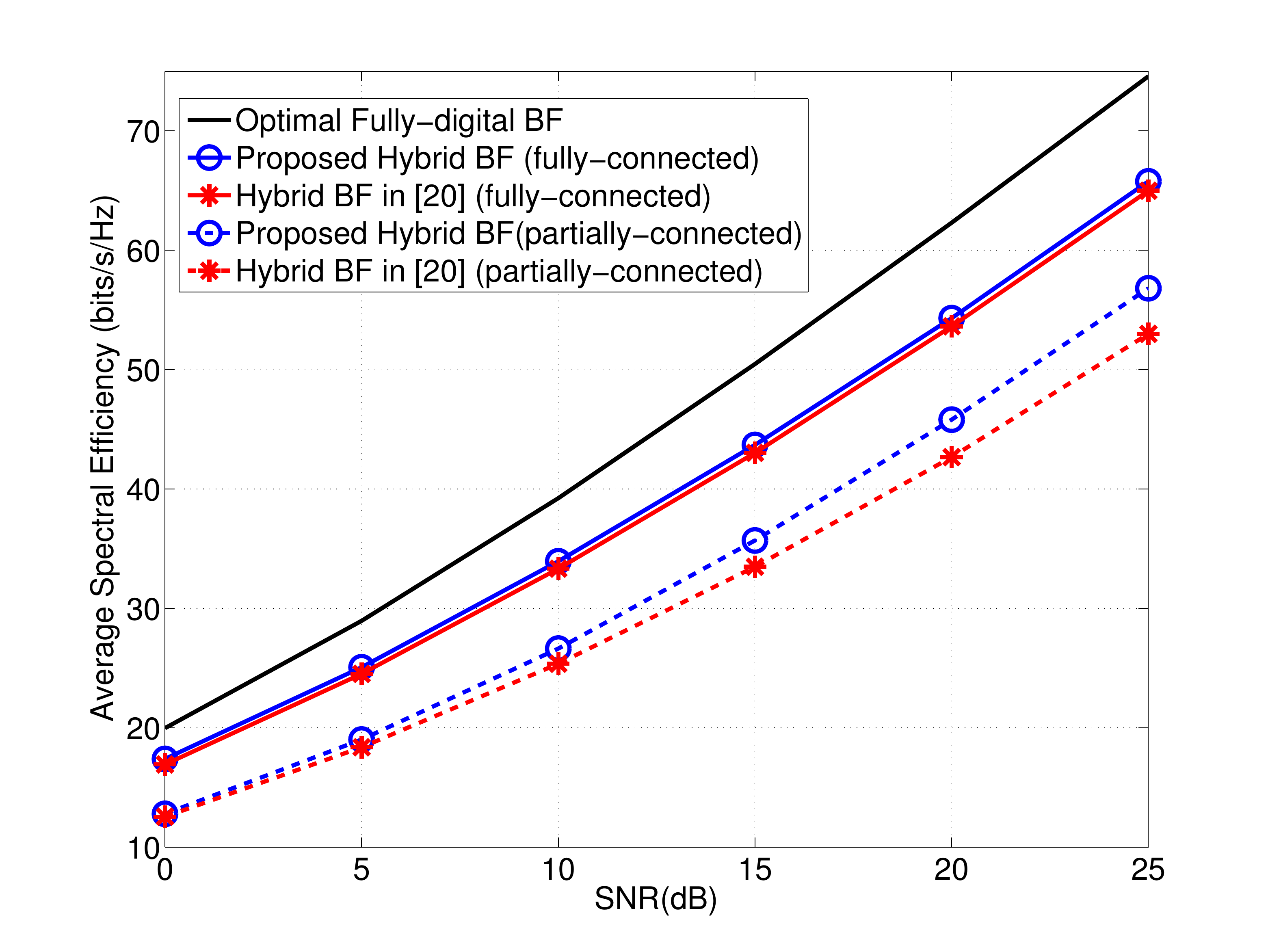}
	\centering
	\caption{Spectral efficiencies versus SNR of different methods for a $64\times 8$ OFDM-based SU-MIMO system in which hybrid architecture with infinite resolution phase shifters is employed at the transmitter and $N_s=8$, $N_\text{RF} =8$ and $K=64$. 
	}
	\label{Fig_VsDyn_Inf}
\end{figure}

The spectral efficiency maximization problem has been recently considered in \cite{park2016dynamic} for the case that the hybrid beamforming architecture is employed only at the transmitter. In order to compare the performance of the proposed scheme with the algorithm in \cite{park2016dynamic}, this experiment considers a $64\times 8$ MIMO system with $K=64$ subcarriers in which a transmitter with $8$ RF chains sends $N_s = 8$ data symbols per subcarrier to a receiver equipped with $8$ RF chains so that the receiver can employ fully-digital combining. Fig.~\ref{Fig_VsDyn_Inf} shows that the proposed algorithm for the partially-connected structure achieves a higher spectral efficiency as compared to the hybrid beamforming design in \cite{park2016dynamic}.
However, for the fully-connected case, the achievable rate of the proposed algorithm is very similar to that of the hybrid beamforming design in \cite{park2016dynamic}. This is because \cite{park2016dynamic} 
already utilizes the fact that
for the fully-connected structure, the analog precoder should be designed according to the average of the covariance matrices of frequency domain channels, $\mathbf{F}_1 = \frac{1}{K} {\sum_{k=1}^{K}}\left({\mathbf{H}[k]}^{H} \mathbf{H}[k]\right)$. In particular, the algorithm in \cite{park2016dynamic} first finds the set of eigenvectors of $\mathbf{F}_1$ corresponding to the $N_\text{RF}$ largest eigenvalues and then sets the columns of analog precoder to the phase of that eigenvectors. It can be seen that without the hybrid constraint, the optimal solution to our analog precoder design problem in \eqref{main_problem_transmitter_RF} is also the eigenvectors of $\mathbf{F}_1$. Therefore, it is expected that the algorithm in \cite{park2016dynamic} approaches the performance of the proposed algorithm for the cases that fully-connected hybrid architecture with infinite-resolution phase shifters is employed at the transmitter. 

%%%%%%%%%%%%%%%%%%%%%%%%%%%%%%%%%%%%
%% Fig. 5
%%%%%%%%%%%%%%%%%%%%%%%%%%%%%%%%%%%%

%%%%%%%%%%%%%%%%%%%%%%%%%%%%%%%%%%%%
%% Fig. 6
%%%%%%%%%%%%%%%%%%%%%%%%%%%%%%%%%%%%
\begin{figure}[t]
	\centering
	\includegraphics[width=0.51\textwidth]{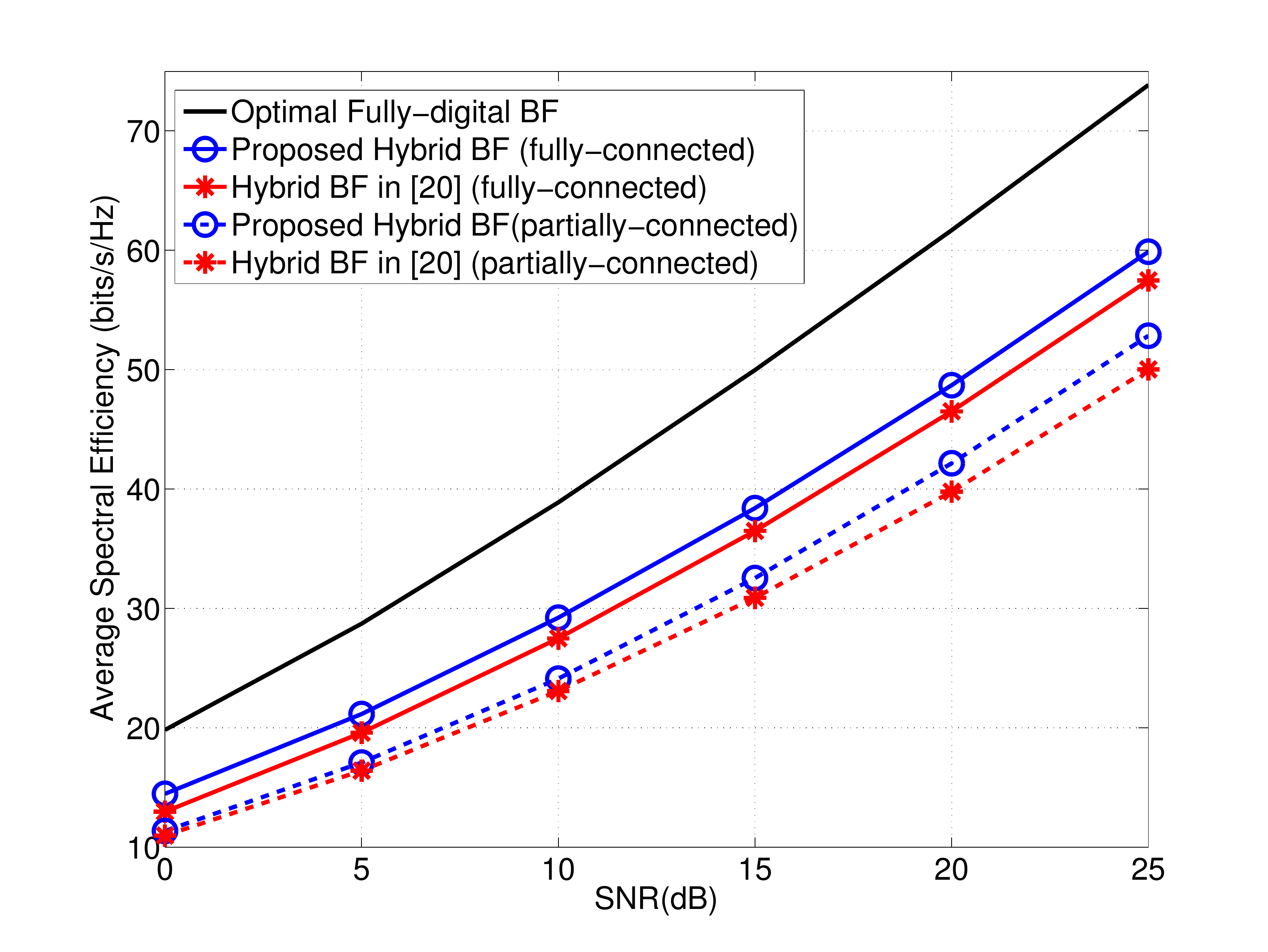}
	\centering
	\caption{Spectral efficiencies versus SNR of different methods for a $64\times 8$ OFDM-based SU-MIMO system in which hybrid architecture with 1-bit resolution phase shifters is employed at the transmitter and $N_s=8$, $N_\text{RF} =8$ and $K=64$.
	}
	\label{Fig_VsDyn_1bit}
\end{figure}

In the next experiment, we consider the same system parameters as the previous experiment except we assume very low-resolution phase-shifters at the transmitter, $b = 1$. Fig.~\ref{Fig_VsDyn_1bit} shows that the proposed algorithm achieves higher performance, i.e., about $1$dB, compared to the algorithm in \cite{park2016dynamic} for such a scenario. Overall, our proposed algorithm for fully-connected structure has two advantages over the algorithm in \cite{park2016dynamic}: i) The proposed algorithm provides hybrid combining design as well as hybrid precoding design while the algorithm in \cite{park2016dynamic} only considers hybrid precoding design with a fully-digital receiver. ii) The performance degradation in hybrid precoding without full degree of freedom, i.e., when the partially connected structure is used or when low-resolution
phase shifters are employed, is greater for the algorithm in \cite{park2016dynamic} as compared to the proposed method.

%%%%%%%%%%%%%%%%%%%%%%%%%%%%%%%%%%%%%%
%%				Hybrid Beamforming Analysis in MU-MISO systems
%%%%%%%%%%%%%%%%%%%%%%%%%%%%%%%%%%%%%%
\subsection{Hybrid Beamforming Analysis in MU-MISO systems}
Finally, we consider a single-cell downlink scenario in which a base station with $64$ antennas serves users from $10$ different clusters in which the users in the same cluster share the same transmit antenna channel response vectors but have different complex channel gains and pathloss. \changet{We assume that many users are randomly and uniformly located in a circular coverage area of radius $R = 0.2$km.} At each time slot, $N_U = 4$ users are randomly scheduled to be served over the entire band of $32$ MHz with $32$ subcarriers. Further, it is assumed that the pathloss for a user with distance $\bar{d}$ to the base station is modeled as $128.1 + 37.6\operatorname{log}_{10}(\bar{d})$. \changet{Note that the  random scheduling of the users implies that there is a possibility that in each time slot some scheduled users 
are from the same cluster, or they can be from 
 different clusters. Scheduling users from the same cluster can be useful for increasing the intended signal power since the analog beamformer can be matched to the similar channel response vectors of those users, but this also implies higher inter-user interference. Therefore, it is not straightforward to see whether scheduling users from the same cluster is beneficial for rate maximization or not. 
We leave the optimization of user scheduling for future work and assume the simple uniformly random scheduling in this paper.}

\begin{figure}[t]
	\centering
	\includegraphics[width=0.51\textwidth]{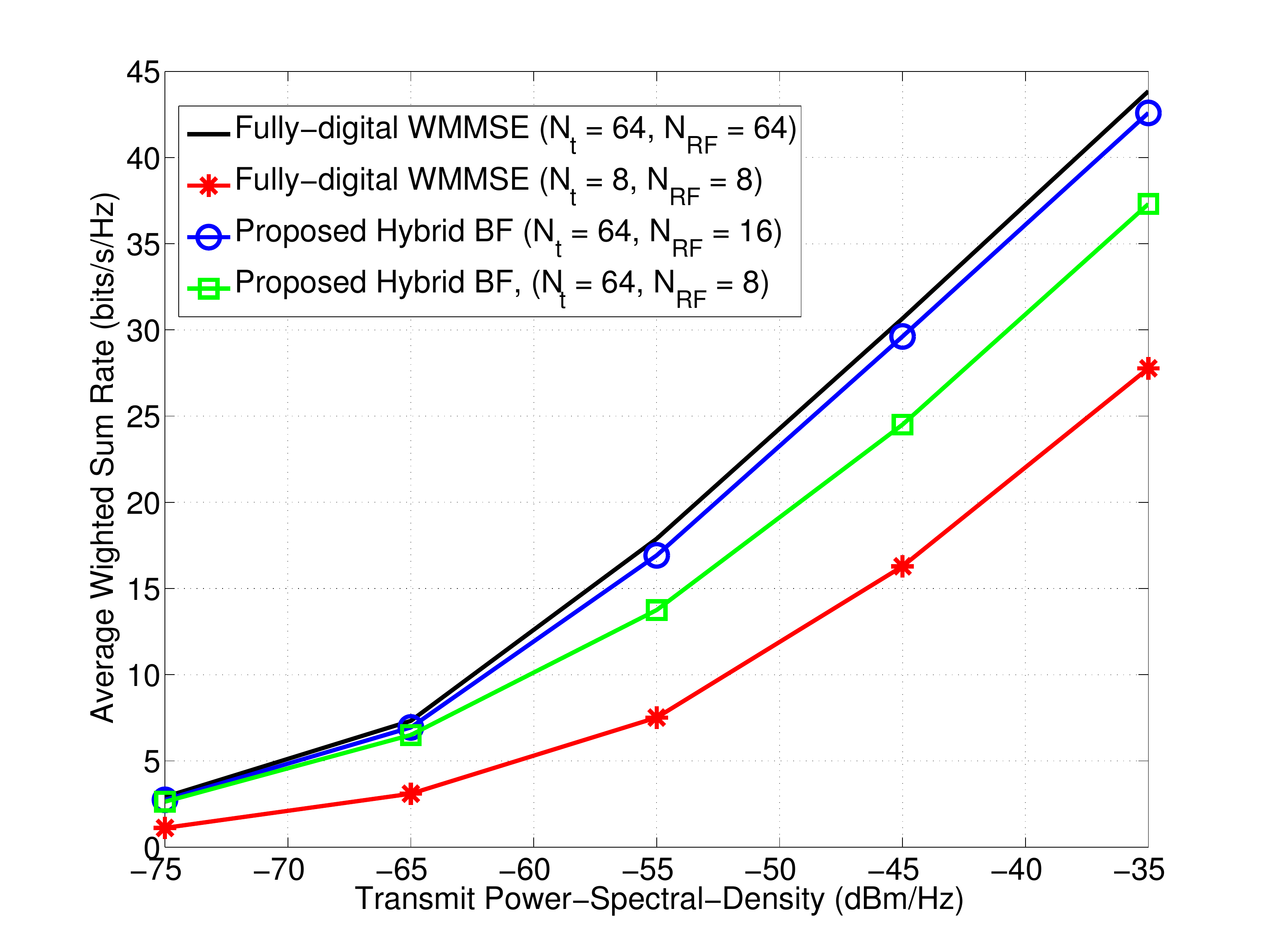}
	\centering
	\caption{Weighted sum rate versus transmit power spectral density for different methods in an OFDM-based MU-MISO system with $N_U = 4$, $N_t = 64$ and $K=32$.
	}
	\label{fig:multiuser_snr}
\end{figure}

\changet{In the first experiment of this part, we assume that the priority weight of each user is set to be proportional to the inverse of the expected rate of that user when the transmitted power spectral density is $-55$ dBm/Hz. Further, the power spectral density of the noise is considered to be $-139$ dBm/Hz. Fig.~\ref{fig:multiuser_snr} plots the average weighted sum rate for the proposed hybrid beamforming algorithm with different number of RF chains and the fully-digital WMMSE algorithm in \cite{shi2011iteratively} for different number of antennas. It can be seen that the proposed hybrid beamforming algorithm with $64$ antennas and $8$ RF chains achieves much higher spectral efficiency as compared to the fully-digital WMMSE beamforming  utilizing  $8$ antennas and $8$ RF chains. This means that using large-scale antenna arrays can be very beneficial even if the number of available RF chians is limited. Fig.~\ref{fig:multiuser_snr} also shows that the proposed hybrid beamforming design with $16$ RF chains approach the performance of the fully-digital WMMSE beamforming with $64$ antennas.}

\begin{figure}[t]
	\centering
	\includegraphics[width=0.51\textwidth]{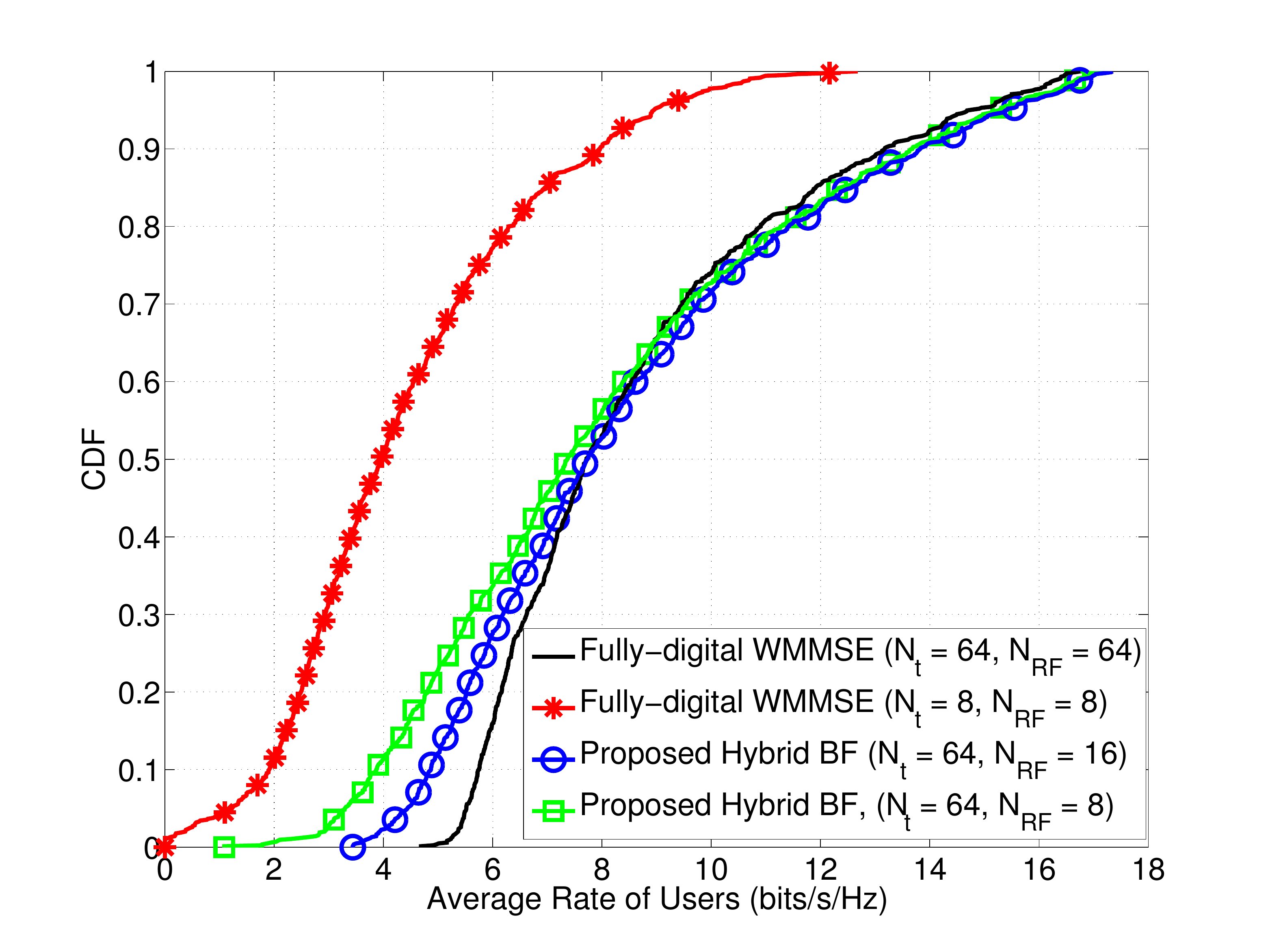}
	\centering
	\caption{Empirical CDF of the average rate of the users for different methods in an OFDM-based MU-MISO system with $N_U = 4$, $N_t = 64$ and $K=32$.
	}
	\label{fig:multiuser_cdf}
\end{figure}

\changet{
In the second experiment, we consider a more practical setup in which the priority weight of each user is adapted over many iterations by setting it to be proportional to the inverse of the experimentally average rate achieved so far. The power spectral densities of the transmitted signals and the noise are set to be $-45$ dBm/Hz and $-139$ dBm/Hz, respectively. The other parameters are similar to the previous experiment. Fig.~\ref{fig:multiuser_cdf} plots the empirical cumulative distribution function (CDF) of the average rate achieved by the cell users.  Fig.~\ref{fig:multiuser_cdf} indicates that the proposed hybrid beamforming strategy with both $N_\text{RF} = 8$ and $N_\text{RF} = 16$ can approximately achieve the same sum rate as the fully-digital WMMSE approach with $N_t = 64$ antennas and $N_\text{RF} = 64$ RF chains. However, it can seen from Fig.~\ref{fig:multiuser_cdf} that the lower $40$-percentile users still achieve better rates in fully-digital WMMSE beamforming scheme while the other higher $60$-percentile users achieve slightly better rates in the proposed hybrid beamforming design. This is because of the fact that the proposed analog precoder is designed for maximizing the sum rate, hence it may favor high rate users at the expense of low rate users.}

%%%%%%%%%%%%%%%%%%%%%%%%%%%%%%%%%%%%%
%%% Fig. 7
%%%%%%%%%%%%%%%%%%%%%%%%%%%%%%%%%%%%%

%%%%%%%%%%%%%%%%%%%%%%%%%%%%%%%%%%%%%
%%% Fig. 8
%%%%%%%%%%%%%%%%%%%%%%%%%%%%%%%%%%%%%

%%%%%%%%%%%%%%%%?%%%%%%%%%%%%%%%%%%%%%
%%% Conclusion
%%%%%%%%%%%%%%%%%%%%%%%%%%%%%%%%%%%%%
\section{Conclusion}
\label{sec:con}
This paper considers hybrid beamforming design for OFDM-based systems with large-scale antenna arrays. We first show that the hybrid beamforming architecture is an appropriate scheme for broadband mmWave systems with frequency-selective channels. In particular, for SU-MIMO systems, we show that the hybrid structure can asymptotically realize the optimal fully-digital beamforming for sufficiently large number of antennas. Then, for practical number of antennas, we propose a unified heuristic algorithm for designing the hybrid precoders and combiners for two well-known architectures:
fully-connected and partially-connected hybrid beamforming. Finally, we generalize the proposed algorithm for designing the hybrid beamforming in the downlink of a MU-MISO system. The simulation results verify that the proposed algorithm can achieve a better performance as compared to the existing methods for both architectures. Further, it is shown that the proposed design for the fully-connected architecture can approach  
the performance of the optimal fully-digital beamforming with \changett{a} reasonable number of RF chains, which is typically much less than the number of antennas.

%%%%%%%%%%%%%%%%%%%%%%%%%%%%%%%%%%%%
%% Refrences
%%%%%%%%%%%%%%%%%%%%%%%%%%%%%%%%%%%%
\bibliographystyle{IEEEtran}
\bibliography{IEEEabrv,refrences}

\begin{IEEEbiography}[{\includegraphics[width=1in,height=1.25in,clip,keepaspectratio]{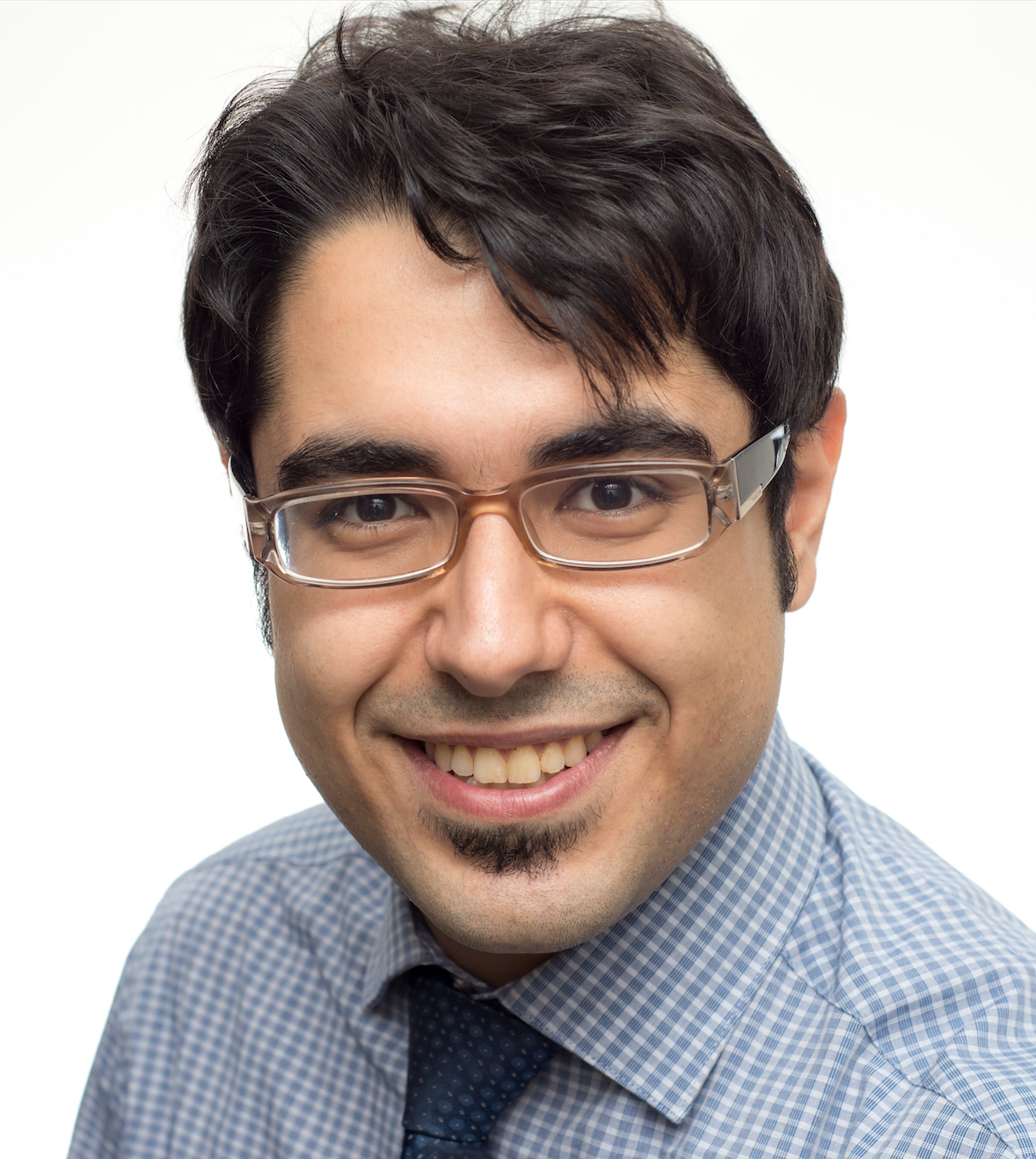}}]{Foad Sohrabi}
(S'13) received his B.A.Sc.\ degree in 2011 from the University of Tehran, Tehran, Iran, and his M.A.Sc.\ degree in 2013 from McMaster University, Hamilton, ON, Canada, both in Electrical and Computer Engineering. Since September 2013, he has been a Ph.D student at University of Toronto, Toronto, ON, Canada. Form July to December 2015, he was a research intern at Bell-Labs, Alcatel-Lucent, in Stuttgart, Germany. His main research interests include MIMO communications, optimization theory, wireless communications, and signal processing.
\end{IEEEbiography}

\begin{IEEEbiography}[{\includegraphics[width=1in,height=1.25in,clip,keepaspectratio]{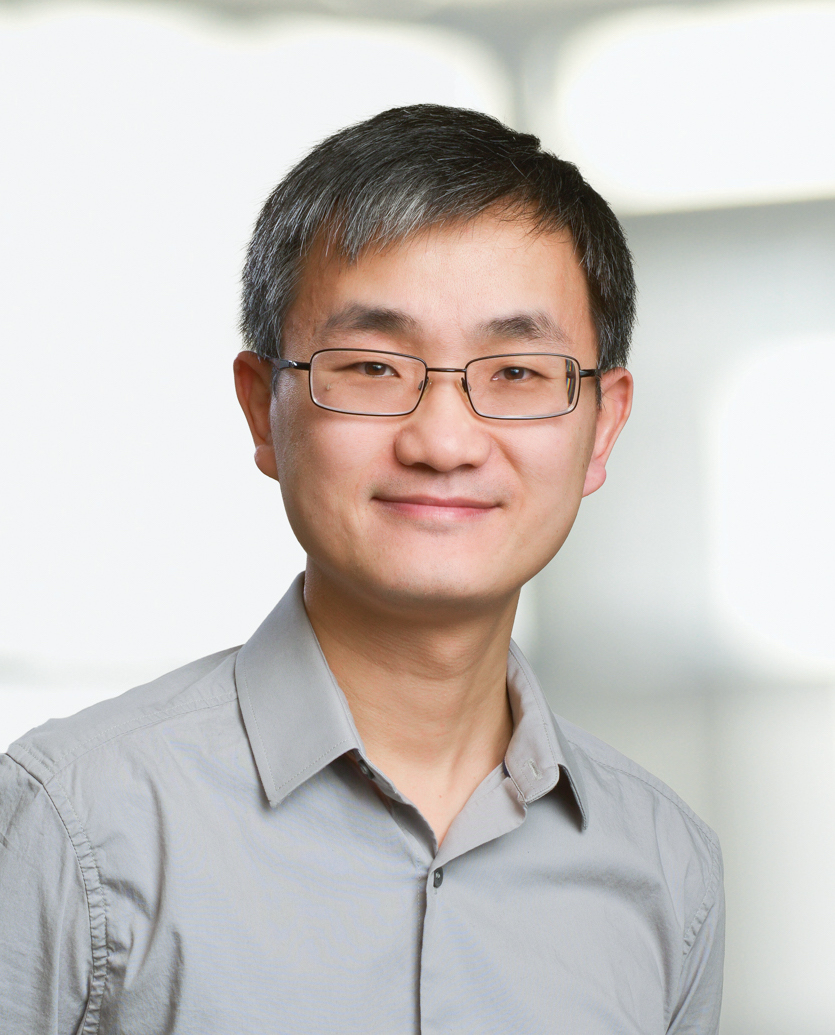}}]{Wei Yu} (S'97-M'02-SM'08-F'14) received the B.A.Sc. degree in Computer Engineering and Mathematics from the University of Waterloo, Waterloo, Ontario, Canada in 1997 and M.S. and Ph.D. degrees in Electrical Engineering from Stanford University, Stanford, CA, in 1998 and 2002, respectively. Since 2002, he has been with the Electrical and Computer Engineering Department at the University of Toronto, Toronto, Ontario, Canada, where he is now Professor and holds a Canada Research Chair (Tier 1) in Information Theory and Wireless Communications. His main research interests include information theory, optimization, wireless communications and broadband access networks.

Prof. Wei Yu currently serves on the IEEE Information Theory Society Board of Governors (2015-17). He serves as the Chair of the Signal Processing for Communications and Networking Technical Committee of the IEEE Signal Processing Society (2017-18). He was an IEEE Communications Society Distinguished Lecturer (2015-16). He served as an Associate Editor for \textsc{IEEE Transactions on Information Theory} (2010-2013), as an Editor for IEEE \textsc{Transactions on Communications} (2009-2011), as an Editor for \textsc{IEEE Transactions on Wireless Communications} (2004-2007), and as a Guest Editor for a number of special issues for the \textsc{IEEE Journal on Selected Areas in Communications} and the \textsc{EURASIP Journal on Applied Signal Processing}. He was a Technical Program co-chair of the IEEE Communication Theory Workshop in 2014, and a Technical Program Committee co-chair of the Communication Theory Symposium at the IEEE International Conference on Communications (ICC) in 2012. Prof. Wei Yu received a Steacie Memorial Fellowship in 2015, an IEEE Communications Society Best Tutorial Paper Award in 2015, an IEEE ICC Best Paper Award in 2013, an IEEE Signal Processing Society Best Paper Award in 2008, the McCharles Prize for Early Career Research Distinction in 2008, the Early Career Teaching Award from the Faculty of Applied Science and Engineering, University of Toronto in 2007, and an Early Researcher Award from Ontario in 2006. He is recognized as a Highly Cited Researcher by Thomson Reuters. He is recognized as a Highly Cited Researcher by Clarivate Analytics.
\end{IEEEbiography}


\end{document}